\documentstyle[amstex,amsfonts,epsfig,12pt]{article}
\begin{document}

\onecolumn

\begin{titlepage}
\begin{center}
\hfill{WATPHYS-TH01/06}
\\ \vspace{1.5cm}
{\LARGE \bf Particles on a Circle in Canonical Lineal Gravity}
\\ \vspace{2cm}
R.B. Mann \footnotemark\footnotetext{email: mann@@avatar.uwaterloo.ca} 
\\
\vspace{1cm}
Dept. of Physics,
University of Waterloo
Waterloo, ONT N2L 3G1, Canada\\
\vspace{2cm}
\vspace{2cm}
\today\\
\end{center}

\begin{abstract}
A description of the canonical formulation of lineal gravity
minimally coupled to N point particles in a  circular
topology is given.  The Hamiltonian 
is found to be equal to the time-rate of change of the extrinsic curvature
multiplied by the proper circumference of the circle.  Exact solutions for pure
gravity and for gravity coupled to a single particle are obtained. 
The presence of a single particle significantly modifies the spacetime evolution
by either slowing down or reversing the cosmological expansion of the circle,
depending upon the choice of parameters.
\end{abstract}
\end{titlepage}\onecolumn

\section{INTRODUCTION}

\bigskip

An increasing amount of attention is being given to the problem of
lower-dimensional self-gravitating systems. \ These systems consist of a
collection of $\ N$ particles mutually interacting through their own mutual
gravitational attraction, along with other specified forces. \ They are used
not only as prototypes for the behaviour of gravity in higher dimensions,
but can also also approximate the behaviour of some physical systems in 3
spatial dimensions, such as the dynamics of stars in a direction orthogonal
to the plane of a highly flattened galaxy, the collisions of flat parallel
domain walls, and the dynamics of cosmic strings. For the many-body case
there has been much work on understanding the fractal behaviour \cite%
{fractal} and ergodic and equipartition properties \cite{yawn} of
non-relativistic one-dimensional self-gravitating systems. Only recently has
this been extended to included relativistic effects \cite{pchak}. \ For the
2-body case there is an exact relativistic solution in 2 spatial dimensions %
\cite{italiangroup}, although it does not have a non-relativistic limit.
Several exact solutions to the 2-body problem have been found in one spatial
dimension \cite{2bd}. These have a non-relativistic limit\thinspace\ \cite%
{OR}, and have been extended to include both cosmological expansion \cite%
{2bdcossh,2bdcoslo}\ and electromagnetic interactions \cite{2bdchglo}. All
solutions obtained so far have been for non-compact spatial dimensions. \ 

\bigskip

The purpose of this paper is to extend to circular topology the $N$-body
problem for relativistic gravity in $(1+1)$ dimensions (i.e. lineal
gravity). \ This is the only compact topology available in one spatial
dimension and it introduces qualitatively new features not present in the
non-compact case. For example, an analogous non-relativistic solution does
not exist (as is the case in higher dimensions). \ This is easily seen by
considering the non-relativistic canonical field equations for a point
source in one spatial dimension%
\begin{eqnarray}
\varphi ^{\prime \prime } &=&m\;\delta (x-z(t))\;  \label{I1} \\
\dot{p} &=&-\varphi ^{\prime }\left( z\right)  \label{I2} \\
\dot{z_{a}} &=&\frac{p}{m}  \label{I3}
\end{eqnarray}%
where the prime refers to the derivative with respect to the spatial
coordinate $x$. The first equation yields a solution for the gravitational
potential $\varphi $ which grows linearly with $x$. \ This is consistent
with the remaining equations, taking $p=z=0$ and $\varphi =\frac{m}{2}\left|
x\right| $, which has vanishing derivative at the origin. For a lineal
topology this is fine, but for a circular topology we must have both $%
\varphi (L)=\varphi (-L)$ and \ $\varphi ^{\prime }(L)=\varphi ^{\prime
}(-L) $ for some $L$, where $2L$ is the circumference of the circle. There
is no solution to these matching conditions unless another point source of
negative mass is introduced. \ \ For any compact smeared source the problem
is the same: the potential grows linearly with increasing distance from the
source and the matching conditions cannot be satisfied for physically
reasonable (i.e. positive mass) sources. \ However in a dynamical spacetime
this problem has a solution since the spacetime can expand or contract in
response to the presence of the source. \ 

In this paper I formulate a general framework for canonical reduction of
lineal gravity minimally coupled to $N$ particles in the presence of a
cosmological constant $\Lambda $. \ This extends previous work done in the
non-compact case \cite{OR} and is somewhat analogous to the reduction of the 
$(2+1)$ dimensional Einstein equations for spatially compact manifolds to a
Hamiltonian system \cite{Moncreif}. \ Choosing the mean curvature to play
the role of time, I find that the Hamiltonian becomes the circumference
functional of the circle. \ After establishing the basic formalism, I solve
the canonical equations for $N=0$ (pure gravity) and $N=1$. \ 

The lineal gravity theory chosen here is one that models 4D general
relativity in that it sets the Ricci scalar equal to the trace of the
stress-energy of prescribed matter fields and sources. This theory
(sometimes referred to as $R=T$ theory) has the property that matter governs
the evolution of spacetime curvature which reciprocally governs the
evolution of matter \cite{r3}. It has a consistent Newtonian limit \cite{r3}%
, a problematic limit in a generic $(1+1)$-dimensional theory of gravity
theory \cite{jchan}. Setting the particle stress-energy to zero, leaving
only the cosmological constant, the theory reduces to Jackiw-Teitelboim (JT)
theory \cite{JT}. \ Since pure gravity in $(1+1)$-dimensions has no dynamics
(the Einstein-Hilbert action is a topological invariant) it is necessary to
include a scalar (dilaton) field in the action \cite{BanksMann}. \ The
particular choice of dilaton coupling in $R=T$ theory is such that the
evolution of the dilaton does not modify the reciprocal gravity/matter
dynamics noted above.

The outline of the paper is as follows. In section 2 the canonical
formulation of the $N$-particle self-gravitating system in $(1+1)$
dimensions is given, and the Hamiltonian is shown to be proportional to the
circumference functional of the circle multiplied by the time-rate-of-change
of the extrinsic curvature.\ In section 3 the equations are solved in the
pure gravity case with cosmological constant using two different methods. \
In section 4 the equations are solved in the single particle case, and then
analyzed using a variety of different choices for the time dependence of the
extrinsic curvature. The last section contains a summary of the work and
some suggestions for further research.

\bigskip

\section{CANONICAL REDUCTION OF THE $N$-PARTICLE SYSTEM}

\bigskip

The canonical reduction of \ the $N$-body problem in $(1+1)$-dimensions with
circular topology has several features in common with that of its
non-compact counterpart \cite{OR,2bd,2bdcoslo}. \ The action integral for
the gravitational fields coupled with $N$ point masses is 
\begin{eqnarray}
I &=&\int d^{2}x\left[ \frac{1}{2\kappa }\sqrt{-g}g^{\mu \nu }\left\{ \Psi
R_{\mu \nu }+\frac{1}{2}\nabla _{\mu }\Psi \nabla _{\nu }\Psi +\frac{1}{2}%
g_{\mu \nu }\Lambda \right\} \right.  \nonumber \\
&&\qquad \left. +\sum_{a}\int d\tau _{a}\left\{ -m_{a}\left( -g_{\mu \nu }(x)%
\frac{dz_{a}^{\mu }}{d\tau _{a}}\frac{dz_{a}^{\nu }}{d\tau _{a}}\right)
^{1/2}\right\} \delta ^{2}(x-z_{a}(\tau _{a}))\right] \;,  \label{act0}
\end{eqnarray}%
where $\Psi $ is the dilaton field, $g_{\mu \nu }$ and $g$ are the metric
and its determinant, $R$ is the Ricci scalar, and $e_{a}$ and $\tau _{a}$
are the charge and the proper time of $a$-th particle, respectively, with $%
\kappa =8\pi G/c^{4}$. The symbol $\nabla _{\mu }$ denotes the covariant
derivative associated with $g_{\mu \nu }$. Here I take the range of $x$ to
be $-L\leq x\leq L$, and the circular topology implies that all fields must
be smooth (or at least $C^{1}$) functions of $x$ with period $2L$, which
implies 
\begin{equation}
f(L)=f(-L)\qquad \text{and}\qquad f^{\prime }(L)=f^{\prime }(-L)=0\;\;.
\label{smooth}
\end{equation}%
for all functions.

\bigskip The field equations derived from the action (\ref{act0}) are 
\begin{eqnarray}
&&  \nonumber \\
&&R-g^{\mu \nu }\nabla _{\mu }\nabla _{\nu }\Psi =0\;,  \label{eq-R} \\
&&\frac{1}{2}\nabla _{\mu }\Psi \nabla _{\nu }\Psi -\frac{1}{4}g_{\mu \nu
}\nabla ^{\lambda }\Psi \nabla _{\lambda }\Psi +g_{\mu \nu }\nabla ^{\lambda
}\nabla _{\lambda }\Psi -\nabla _{\mu }\nabla _{\nu }\Psi =\kappa T_{\mu \nu
}+\frac{1}{2}g_{\mu \nu }\Lambda \;,  \label{eq-Psi} \\
&&m_{a}\left[ \frac{d}{d\tau _{a}}\left\{ g_{\mu \nu }(z_{a})\frac{%
dz_{a}^{\nu }}{d\tau _{a}}\right\} -\frac{1}{2}g_{\nu \lambda ,\mu }(z_{a})%
\frac{dz_{a}^{\nu }}{d\tau _{a}}\frac{dz_{a}^{\lambda }}{d\tau _{a}}\right]
=0\;,  \label{eq-z}
\end{eqnarray}%
where 
\begin{equation}
T_{\mu \nu }=\sum_{a}m_{a}\int d\tau _{a}\frac{1}{\sqrt{-g}}g_{\mu \sigma
}g_{\nu \rho }\frac{dz_{a}^{\sigma }}{d\tau _{a}}\frac{dz_{a}^{\rho }}{d\tau
_{a}}\delta ^{2}(x-z_{a}(\tau _{a}))\;,  \label{eq-pointstress}
\end{equation}%
is the stress-energy due to the point masses. Conservation of $T_{\mu \nu }$
is ensured by eq.(\ref{eq-Psi}). Note that insertion of the trace of eq.(\ref%
{eq-Psi}) into (\ref{eq-R}) yields 
\begin{equation}
R-\Lambda =\kappa T_{\;\;\mu }^{\mu }\;.  \label{RT}
\end{equation}%
Eqs. (\ref{eq-z}) and (\ref{RT}) form a closed sytem of equations for
gravity and matter.

To canonically reduce this system, write the metric in the form 
\begin{equation}
ds^{2}=-N_{0}^{2}dt^{2}+\gamma \left( dx+\frac{N_{1}}{\gamma }dt\right)
^{2}\;,  \label{lineel}
\end{equation}%
so that $\gamma =g_{11},N_{0}=(-g^{00})^{-1/2}$ and $N_{1}=g_{10}$. \
Decomposing the scalar curvature in terms of the extrinsic curvature $K$ via 
\begin{equation}
\sqrt{-g}R=-2\partial _{0}(\sqrt{\gamma }K)+2\partial _{1}[(N_{1}K-\partial
_{1}N_{0})/\sqrt{\gamma }]\;,  \label{extK}
\end{equation}%
where 
\begin{equation}
K=(2N_{0}\gamma )^{-1}(2\partial _{1}N_{1}-\gamma ^{-1}N_{1}\partial
_{1}\gamma -\partial _{0}\gamma )  \label{Kext}
\end{equation}%
yields for the action (\ref{act0}) 
\begin{equation}
I=\int d^{2}x\left\{ \sum_{a}p_{a}\dot{z}_{a}\delta (x-z_{a}(t))+\pi \dot{%
\gamma}+\Pi \dot{\Psi}+N_{0}R^{0}+N_{1}R^{1}\right\}  \label{act2}
\end{equation}%
where $\pi $ and $\Pi $ are conjugate momenta to $\gamma $ and $\Psi $,
respectively and $p_{a}$ is the momentum conjugate to the coordinate $z_{a}$%
. The constraints are 
\begin{eqnarray}
R^{0} &=&-\kappa \sqrt{\gamma }\gamma \pi ^{2}+2\kappa \sqrt{\gamma }\pi \Pi
+\frac{1}{4\kappa \sqrt{\gamma }}(\Psi ^{\prime })^{2}-\frac{1}{\kappa }%
\left( \frac{\Psi ^{\prime }}{\sqrt{\gamma }}\right) ^{\prime }-\frac{1}{2}%
\sqrt{\gamma }(-\frac{\Lambda }{\kappa })  \nonumber \\
&&-\sum_{a}\sqrt{\frac{p_{a}^{2}}{\gamma }+m_{a}^{2}}\;\delta (x-z_{a}(t))\;,
\label{R0} \\
R^{1} &=&\frac{\gamma ^{\prime }}{\gamma }\pi -\frac{1}{\gamma }\Pi \Psi
^{\prime }+2\pi ^{\prime }+\sum_{a}\frac{p_{a}}{\gamma }\delta
(x-z_{a}(t))\;,  \label{R1}
\end{eqnarray}%
with the symbols $(\;\dot{}\;)$ and $(\;^{\prime }\;)$ denoting $\partial
_{0}$ and $\partial _{1}$, respectively.

From the action (\ref{act2}) the set of field equations is%
\begin{eqnarray}
\dot{\pi} &+&N_{0}\left\{ \frac{3\kappa }{2}\sqrt{\gamma }\pi ^{2}-\frac{%
\kappa }{\sqrt{\gamma }}\pi \Pi +\frac{1}{8\kappa \sqrt{\gamma }\gamma }%
(\Psi ^{\prime })^{2}-\frac{1}{4\sqrt{\gamma }}\frac{\Lambda }{\kappa }%
\right.  \nonumber \\
&&\makebox[10em]{}\left. -\sum_{a}\frac{p_{a}^{2}}{2\gamma ^{2}\sqrt{\frac{%
p_{a}^{2}}{\gamma }+m_{a}^{2}}}\;\delta (x-z_{a}(t))\right\}  \nonumber \\
&+&N_{1}\left\{ -\frac{1}{\gamma ^{2}}\Pi \Psi ^{\prime }+\frac{\pi ^{\prime
}}{\gamma }+\sum_{a}\frac{p_{a}}{\gamma ^{2}}\;\delta (x-z_{a}(t))\right\}
+N_{0}^{\prime }\frac{1}{2\kappa \sqrt{\gamma }\gamma }\Psi ^{\prime
}+N_{1}^{\prime }\frac{\pi }{\gamma }=0\;,  \label{e-pidot}
\end{eqnarray}%
\begin{eqnarray}
&&\dot{\gamma}-N_{0}(2\kappa \sqrt{\gamma }\gamma \pi -2\kappa \sqrt{\gamma }%
\Pi )+N_{1}\frac{\gamma ^{\prime }}{\gamma }-2N_{1}^{\prime }=0\;,
\label{e-gamma} \\
&&R^{0}=0\;,  \label{e-R0} \\
&&R^{1}=0\;,  \label{e-R1} \\
&&\dot{\Pi}+\partial _{1}(-\frac{1}{\gamma }N_{1}\Pi +\frac{1}{2\kappa \sqrt{%
\gamma }}N_{0}\Psi ^{\prime }+\frac{1}{\kappa \sqrt{\gamma }}N_{0}^{\prime
})=0\;,  \label{e-Pi} \\
&&\dot{\Psi}+N_{0}(2\kappa \sqrt{\gamma }\pi )-N_{1}(\frac{1}{\gamma }\Psi
^{\prime })=0\;,  \label{e-Psi} \\
&&\dot{p}_{a}+\frac{\partial N_{0}}{\partial z_{a}}\sqrt{\frac{p_{a}^{2}}{%
\gamma }+m_{a}^{2}}-\frac{N_{0}}{2\sqrt{\frac{p_{a}^{2}}{\gamma }+m_{a}^{2}}}%
\frac{p_{a}^{2}}{\gamma ^{2}}\frac{\partial \gamma }{\partial z_{a}} 
\nonumber \\
&&\makebox[2em]{}-\frac{\partial N_{1}}{\partial z_{a}}\frac{p_{a}}{\gamma }%
+N_{1}\frac{p_{a}}{\gamma ^{2}}\frac{\partial \gamma }{\partial z_{a}}=0\;,
\label{e-p} \\
&&\dot{z_{a}}-N_{0}\frac{\frac{p_{a}}{\gamma }}{\sqrt{\frac{p_{a}^{2}}{%
\gamma }+m_{a}^{2}}}+\frac{N_{1}}{\gamma }=0\;.  \label{e-z}
\end{eqnarray}%
In equations (\ref{e-p}) and (\ref{e-z}), all metric components ($N_{0}$, $%
N_{1}$, $\gamma $) are evaluated at the point $x=z_{a}$ and 
\[
\frac{\partial f}{\partial z_{a}}\equiv \left. \frac{\partial f(x)}{\partial
x}\right| _{x=z_{a}}\;. 
\]%
The quantities $N_{0}$ and $N_{1}$ are Lagrange multipliers which yield the
constraint equations (\ref{e-R0}) and (\ref{e-R1}). The above set of
equations can be proved to be equivalent to the set of equations (\ref{eq-R}%
), (\ref{eq-Psi}) and (\ref{eq-z}).

\bigskip

Using eq. (\ref{e-gamma}), the extrinsic curvature (\ref{Kext}) can be
written in the form 
\begin{equation}
K=\sqrt{\gamma }\kappa (\pi -\Pi /\gamma )  \label{tauK}
\end{equation}%
which can be rewritten as $\qquad $%
\begin{equation}
\pi =\frac{\Pi }{\gamma }+\frac{\tau }{\sqrt{\gamma }\kappa }  \label{tau}
\end{equation}%
where the mean extrinsic curvature $K$ (which is the total extrinsic
curvature in one spatial dimension) is taken to be a time coordinate $\tau
\left( t\right) $ . Choosing a slicing so that each hypersurface has
constant mean extrinsic curvature yields\ \ $\tau ^{\prime }=0\;$.

Inserting (\ref{tau}) into the system (\ref{e-pidot}--\ref{e-z}) and
simplifying with the constraint equations (\ref{e-R0},\ref{e-R1}) yields the
following set of equations to be solved in sequence: 
\begin{subequations}
\label{eqsgamprime}
\begin{eqnarray}
&&  \nonumber \\
&&2\Pi ^{\prime }-\Pi \left( \Psi +\ln \gamma \right) ^{\prime
}+\sum_{a}p_{a}\delta (x-z_{a})=0\;  \label{eqg1} \\
&&\sqrt{\gamma }\left( \frac{1}{\sqrt{\gamma }}\Psi ^{\prime }\right)
^{\prime }-\frac{(\Psi ^{\prime })^{2}}{4}-(\kappa \Pi )^{2}+\gamma \left(
\tau ^{2}-\frac{\Lambda }{2}\right) +\kappa \sum_{a}\sqrt{p_{a}^{2}+\gamma
m_{a}^{2}}\;\delta (x-z_{a})=0  \label{eqg2} \\
&&\sqrt{\gamma }\left( \frac{1}{\sqrt{\gamma }}N_{0}^{\prime }\right)
^{\prime }=\overset{\cdot }{\tau }\gamma +N_{0}\left\{ (\tau ^{2}-\Lambda
/2)\gamma +\kappa \sum_{a}\frac{\gamma m_{a}^{2}}{2\sqrt{p_{a}^{2}+\gamma
m_{a}^{2}}}\;\delta (x-z_{a}(x^{0}))\right\}  \label{eqg3} \\
&&N_{1}^{\prime }=\dot{\gamma}/2-\gamma \tau N_{0}+N_{1}\frac{\gamma
^{\prime }}{2\gamma }  \label{eqg4} \\
&&\dot{\Pi}+\partial _{1}(-\frac{1}{\gamma }N_{1}\Pi +\frac{1}{2\kappa \sqrt{%
\gamma }}N_{0}\Psi ^{\prime }+\frac{1}{\kappa \sqrt{\gamma }}N_{0}^{\prime
})=0  \label{eqg5} \\
&&\dot{\Psi}+2N_{0}\left( \kappa \frac{\Pi }{\sqrt{\gamma }}+\tau \right)
-N_{1}(\frac{1}{\gamma }\Psi ^{\prime })=0  \label{eqg6} \\
&&\dot{p}_{a}+\frac{\partial N_{0}}{\partial z_{a}}\sqrt{\frac{p_{a}^{2}}{%
\gamma }+m_{a}^{2}}-\frac{N_{0}}{2\sqrt{\frac{p_{a}^{2}}{\gamma }+m_{a}^{2}}}%
\frac{p_{a}^{2}}{\gamma ^{2}}\frac{\partial \gamma }{\partial z_{a}}-\frac{%
\partial N_{1}}{\partial z_{a}}\frac{p_{a}}{\gamma }+N_{1}\frac{p_{a}}{%
\gamma ^{2}}\frac{\partial \gamma }{\partial z_{a}}=0  \label{eqg7} \\
&&\dot{z_{a}}-N_{0}\frac{\frac{p_{a}}{\gamma }}{\sqrt{\frac{p_{a}^{2}}{%
\gamma }+m_{a}^{2}}}+\frac{N_{1}}{\gamma }=0\;\;.  \label{eqg8}
\end{eqnarray}

The remaining step is to identify the Hamiltonian. \ Eliminating the
constraints, the action (\ref{act2}) reduces upon insertion of (\ref{tau})
to 
\end{subequations}
\begin{eqnarray}
I &=&\int d^{2}x\left\{ \sum_{a}p_{a}\dot{z}_{a}\delta (x-z_{a}(t))+\left( 
\frac{\Pi }{\gamma }+\frac{\tau }{\sqrt{\gamma }\kappa }\right) \dot{\gamma}%
+\Pi \dot{\Psi}\right\}  \nonumber \\
&=&\int d^{2}x\left\{ \sum_{a}p_{a}\dot{z}_{a}\delta (x-z_{a}(t))+\Pi \left( 
\frac{\dot{\gamma}}{\gamma }+\dot{\Psi}\right) -\frac{2}{\kappa }\dot{\tau}%
\sqrt{\gamma }\right\}  \nonumber \\
&=&\int d^{2}x\left\{ \sum_{a}p_{a}\dot{z}_{a}\delta (x-z_{a})+\Pi \frac{%
\partial }{\partial t}\left( \Psi +\ln \gamma \right) -{\cal H}\right\} \;,
\label{act3}
\end{eqnarray}%
where $z_{a}=z_{a}(x^{0})$ is a function of the coordinate $x^{0}\ $and
boundary terms have been dropped. The quantities $\Psi ,\Pi $ and $\gamma $
are functions of $z_{a}$ and $p_{a}$, determined by solving the constraints (%
\ref{eqg1},\ref{eqg2}).

The reduced action has the general form $\int dt\left[ P\frac{dQ}{dt}-H%
\right] $, thus allowing the reduced Hamiltonian for the system of particles
to be identified as 
\begin{equation}
H=\int dx{\cal H=}\frac{2\dot{\tau}}{\kappa }\int dx\sqrt{\gamma }{\cal \;}
\label{ham1}
\end{equation}%
which is the circumference functional of the circle when $\dot{\tau}$ is
constant. \ In contrast to the line topology \cite{marco},\ the Hamiltonian
explicitly depends on the time and so it is not conserved by the evolution.
\ As in $(2+1)$ dimensions \cite{Moncreif}, the dynamics is that of a
time-dependent system, with the time-dependence corresponding to the
time-varying circumference of the circle of constant mean extrinsic
curvature.\ Since any metric on a circle is globally conformal to a flat
metric, the\ spatial metric can be chosen so that $\gamma =\gamma (t)$ ,
i.e. \ $\gamma ^{\prime }=0\;$. \ By choosing the time so that $\dot{\tau}%
\sqrt{\gamma }$ is then constant, the Hamiltonian will be time-independent.

When $\gamma ^{\prime }=0$ the equations simplify to 
\begin{subequations}
\label{final}
\begin{eqnarray}
&&  \nonumber \\
&&2\Pi ^{\prime }-\Pi \Psi ^{\prime }+\sum_{a}p_{a}\delta
(x-z_{a}(x^{0}))=0\;  \label{fin1a} \\
&&\Psi ^{\prime \prime }-\frac{1}{4}(\Psi ^{\prime })^{2}-(\kappa \Pi
)^{2}+\gamma \left( \tau ^{2}-\Lambda /2\right) +\kappa \sum_{a}\sqrt{%
p_{a}^{2}+\gamma m_{a}^{2}}\;\delta (x-z_{a}(x^{0}))=0  \label{fin2} \\
&&N_{0}^{\prime \prime }=\overset{\cdot }{\tau }\gamma +N_{0}\left\{ (\tau
^{2}-\Lambda /2)\gamma +\kappa \sum_{a}\frac{\gamma m_{a}^{2}}{2\sqrt{%
p_{a}^{2}+\gamma m_{a}^{2}}}\;\delta (x-z_{a}(x^{0}))\right\}  \label{fin3}
\\
&&N_{1}^{\prime }=\dot{\gamma}/2-\gamma \tau N_{0}  \label{fin4} \\
&&\dot{\Pi}+\partial _{1}(-\frac{1}{\gamma }N_{1}\Pi +\frac{1}{2\kappa \sqrt{%
\gamma }}N_{0}\Psi ^{\prime }+\frac{1}{\kappa \sqrt{\gamma }}N_{0}^{\prime
})=0  \label{fin5} \\
&&\dot{\Psi}+2N_{0}\left( \kappa \frac{\Pi }{\sqrt{\gamma }}+\tau \right)
-N_{1}(\frac{1}{\gamma }\Psi ^{\prime })=0  \label{fin6} \\
&&\dot{p}_{a}+\frac{\partial N_{0}}{\partial z_{a}}\sqrt{\frac{p_{a}^{2}}{%
\gamma }+m_{a}^{2}}-\frac{N_{0}}{2\sqrt{\frac{p_{a}^{2}}{\gamma }+m_{a}^{2}}}%
\frac{p_{a}^{2}}{\gamma ^{2}}\frac{\partial \gamma }{\partial z_{a}}-\frac{%
\partial N_{1}}{\partial z_{a}}\frac{p_{a}}{\gamma }+N_{1}\frac{p_{a}}{%
\gamma ^{2}}\frac{\partial \gamma }{\partial z_{a}}=0  \label{fin7} \\
&&\dot{z_{a}}-N_{0}\frac{\frac{p_{a}}{\gamma }}{\sqrt{\frac{p_{a}^{2}}{%
\gamma }+m_{a}^{2}}}+\frac{N_{1}}{\gamma }=0\;\;.  \label{fin8}
\end{eqnarray}%
and it is these equations that I shall solve for the $0$ and single particle
cases in subsequent sections.

\section{The 0-particle case (pure gravity)}

\bigskip

When there are no particles set $p=0$, $m=0$, and $z=0$. Equation (\ref%
{fin1a}) is trivially solved to yield 
\end{subequations}
\begin{equation}
\Pi =\Pi _{0}(t)\exp [\Psi /2]  \label{fin1sol}
\end{equation}%
The constraint (\ref{fin2}) and equations (\ref{fin3}) to (\ref{fin6})\ \
then reduce to\label{opartset} 
\begin{eqnarray}
&&  \nonumber \\
\Psi ^{\prime \prime }-\frac{1}{4}(\Psi ^{\prime })^{2}-(\kappa \Pi
_{0}(t))^{2}\exp [\Psi ]+\gamma \left( \tau ^{2}-\Lambda /2\right) = &&0
\label{N0p1} \\
N_{0}^{\prime \prime }-\overset{\cdot }{\tau }\gamma -N_{0}(\tau
^{2}-\Lambda /2)\gamma = &&0  \label{N0p2} \\
N_{1}^{\prime }-\dot{\gamma}/2+\gamma \tau N_{0}= &&0  \label{n0p3} \\
\dot{\Pi}+\partial _{1}(-\frac{1}{\gamma }N_{1}\Pi +\frac{1}{2\kappa \sqrt{%
\gamma }}N_{0}\Psi ^{\prime }+\frac{1}{\kappa \sqrt{\gamma }}N_{0}^{\prime
})= &&0  \label{n0p4} \\
\dot{\Psi}+2N_{0}\left( \kappa \frac{\Pi }{\sqrt{\gamma }}+\tau \right)
-N_{1}(\frac{1}{\gamma }\Psi ^{\prime })= &&0  \label{n0p5}
\end{eqnarray}%
which I will proceed two solve in two different ways.

\subsection{$N_{0}^{2}=1$ Solutions}

Since $N_{0}$ must be periodic ($N_{0}(L)=N_{0}(-L)$), the simplest way to
solve (\ref{N0p2})\ \ is to set \ $N_{0}^{2}=1$ and then solve for $\tau $,
choosing it to be an increasing function of $t$. Setting $N_{1}=0$, it is
then straightforward to solve (\ref{n0p3})\ for $\gamma $. \ 

\subsubsection{$\Lambda <0$}

Setting $N_{0}(t)=-1$, \ the solution of \ (\ref{N0p2}) yields 
\begin{equation}
\tau =\frac{1}{\ell }\tan \left( \frac{t}{\ell }\right)  \label{adstau}
\end{equation}%
for $\tau $, where $\ell ^{2}=2/|\Lambda |$. Setting $N_{1}=0$, it is
straightforward to solve (\ref{fin4})\ for $\gamma $: 
\begin{equation}
\gamma =\cos ^{2}\left( \frac{t}{\ell }\right) .  \label{adsgam}
\end{equation}%
The metric is then 
\begin{equation}
ds^{2}=-dt^{2}+\cos ^{2}\left( \frac{t}{\ell }\right) dx^{2}  \label{adsmet}
\end{equation}%
which is the metric for ($1+1)$ AdS spacetime, where $x$ is periodic with
period $2L$. The topology is that of a circle with zero initial radius at $%
t=-\frac{\pi \ell }{2}$ which then expands to a maximum and then recollapses
to zero radius after a finite amount $\pi \ell $ of proper time. \ The
worldsheet is a Lorentzian 2-sphere. The Hamiltonian 
\begin{equation}
H=\frac{4L}{\kappa \ell ^{2}}\sec \left( \frac{t}{\ell }\right) {\cal \;\;}
\label{adsHam}
\end{equation}%
and is both time-dependent and unbounded.

A simple solution to (\ref{N0p1})\ is to take 
\[
\Psi =\Psi (t)\text{ }\Rightarrow \Pi =\pm \frac{1}{\kappa \ell } 
\]%
which trivially solves (\ref{fin1a}) and (\ref{fin5}). \ One then has from (%
\ref{n0p5})

\begin{equation}
\dot{\Psi}-\frac{2}{\ell }\left( \frac{\pm 1+\sin \left( t/\ell \right) }{%
\cos \left( t/\ell \right) }\right) =0\Rightarrow \Psi =-2\ln \left( 1\mp
\sin \left( t/\ell \right) \right)  \label{adspsi}
\end{equation}%
as the two possible solutions for $\Psi (t)$, with the last two equations (%
\ref{fin7}) and (\ref{fin8}) trivial. There are singularities in $\Psi $\ at 
$t=\pm \frac{\pi \ell }{2}$ .

Of course there are other ways to solve the remaining equations. One can
take the maximally extended solution, in which 
\begin{equation}
ds^{2}=-\cosh ^{2}\left( x/\ell \right) \,dt^{2}+dx^{2}  \label{adsmetmax}
\end{equation}%
so that $\gamma =1$. From (\ref{N0p2}) we see that $\tau $ must be constant,
and from (\ref{fin4}) we see that $\tau =0$, and so cannot serve as a time
coordinate. Furthermore, this solution is not periodic ($N_{0}(L)\neq
N_{0}(-L)$) \ and so must be rejected.

\subsubsection{$\Lambda =0$}

Now the solution is given by

\begin{equation}
\tau =-1/t\text{\ and \ }\gamma =\left( t/\ell \right) ^{2}\text{.}
\label{taugamflat}
\end{equation}%
which has the metric 
\begin{equation}
ds^{2}=-dt^{2}+\left( t/\ell \right) ^{2}\,dx^{2}  \label{flatcirmet}
\end{equation}%
where $x$ is periodic with period $2L$, and $\ell $ is an arbitrary
constant. The topology is that of a circle of zero initial radius, whose
radius expands linearly with increasing proper time. The worldsheet is a
cone whose apex is at $t=0$. The Hamiltonian 
\begin{equation}
H=\frac{4L}{\kappa \ell t}{\cal \;\;}  \label{flatH}
\end{equation}%
and is again unbounded.

The solution of (\ref{N0p1})\ is 
\begin{equation}
\Psi =\Psi _{0}\text{ \ and \ }\Pi =\pm \frac{1}{\kappa \ell }\text{ \ \ \ }
\label{psiflatsol}
\end{equation}%
where $\Psi _{0}$ is an arbitrary constant.

Locally the spacetime (\ref{flatcirmet}) is flat and can be described by
coordinates $(T,X)$, where 
\[
t=\sqrt{T^{2}-X^{2}}\text{ \ \ \ \ \ }x=\ell \tanh ^{-1}\left( \frac{X}{T}%
\right) 
\]%
yielding the metric $ds^{2}=-dT^{2}+dX^{2}$. If the coordinate $x$ is
unwrapped, the spacetime describes the upper quandrant of Minkowski
spacetime $\left| X\right| <T$, bounded by the lightcone $T=\pm X$. \
However the periodicity of $x$ restricts the spacetime to a cone lying in
the interior of this lightcone, and the spacetime cannot be extended beyond
this \ 

\subsubsection{\protect\bigskip $\Lambda >0$}

There are three distinct classes of solutions in this case. Throughout I set 
$\ell ^{2}=2/\Lambda $.

\paragraph{The Candlestick}

\bigskip One solution is given by

\begin{equation}
\tau =\tanh \left( t/\ell \right) /\ell \text{ \ and \ }\gamma =\cosh
^{2}\left( t/\ell \right) .  \label{tausol1}
\end{equation}%
so that the metric is 
\begin{equation}
ds^{2}=-dt^{2}+\cosh ^{2}\left( t/\ell \right) \,dx^{2}  \label{dsmet1}
\end{equation}%
where $x$ is periodic with period $2L=2N\pi \ell $ (see below). The topology
is that of a circle with large initial radius that exponentially shrinks
with proper time to a minimal value and then expands again \ -- the
worldsheet is like a candlestick. The Hamiltonian is 
\begin{equation}
H=\frac{4N\pi }{\kappa \ell }\sec \text{h}\left( \frac{t}{\ell }\right) 
{\cal \;\;}  \label{candleH}
\end{equation}%
and is now bounded. The total energy is $\int_{-\infty }^{\infty }Hdt=\frac{%
4N}{\kappa }\pi ^{2}$, and is discretized in units of \ $4\pi ^{2}/\kappa $.
\ 

The solution of (\ref{N0p1})\ is then 
\begin{equation}
\Psi =-2\;\ln \left\{ \beta (t)+\cos (x/\ell )\right\} +\ln \left( \frac{%
1-\beta ^{2}}{\left( \kappa \ell \Pi _{0}(t)\right) ^{2}}\right) \text{ }
\label{dspsi}
\end{equation}%
which is periodic in $x$, provided $L=N\pi \ell $ \ -- the length of the
circle is no longer arbitrary, but is a fixed multiple of the inverse
cosmological constant. Here 
\begin{equation}
\beta (t)={\mbox{sech}}(t/\ell )\text{ \ \ \ and \ \ }\Pi (t)=\frac{\sinh
(t/\ell )}{\left( 1+\cosh (t/\ell )\cos (x/\ell )\right) \kappa \ell }\text{%
\ }  \label{betapsi1}
\end{equation}%
\ so that $\Pi _{0}(t)=\sinh (t/\ell )/(\kappa \ell )$. \ An alternate way
of writing (\ref{dspsi}) is therefore 
\begin{equation}
\Psi =-2\;\ln \left\{ \cosh (t/\ell )\cos (x/\ell )+1\right\} \text{ \ .}
\label{dspsi2}
\end{equation}

\paragraph{\protect\bigskip The Dish}

An alternate solution to (\ref{N0p2}) is 
\begin{equation}
\tau =-\coth \left( t/\ell \right) /\ell \text{ \ and \ }\gamma =\sinh
{}^{2}\left( t/\ell \right)  \label{tausol2}
\end{equation}%
where now the metric is 
\begin{equation}
ds^{2}=-dt^{2}+\sinh ^{2}\left( t/\ell \right) \,dx^{2}\text{ \ .}
\label{dsmet2}
\end{equation}%
The topology is that of a circle with zero initial radius that exponentially
expands with increasing proper time, yielding a worldsheet like a bowl or
dish. \ The Hamiltonian is now%
\begin{equation}
H=\frac{4L}{\kappa \ell ^{2}}\csc \text{h}\left( \frac{t}{\ell }\right) 
{\cal \;\;}  \label{dishH}
\end{equation}%
and diverges at $t=0.$ \ 

The solution of (\ref{N0p1})\ is 
\begin{equation}
\Psi =-2\;\ln \left\{ \cosh (t/\ell )+1\right\} \text{ \ and \ }\Pi =\frac{1%
}{\kappa \ell }\text{ \ \ \ \ .}  \label{psidish}
\end{equation}%
and the periodicity of $x$ is arbitrary. \ 

\paragraph{The Trumpet}

\bigskip The third type of solution is

\begin{equation}
\tau =1/\ell \ \ \ \text{and \ \ \ }\gamma =\exp \left( 2t/\ell \right) .
\label{tausoltrump}
\end{equation}%
with metric 
\begin{equation}
ds^{2}=-dt^{2}+\exp \left( 2t/\ell \right) \,dx^{2}\text{ \ \ .}
\label{dsmet3}
\end{equation}%
The topology is that of a circle with infinitesimally small initial radius
that exponentially expands with increasing proper time: the worldsheet is
like a trumpet. In this case the Hamiltonian vanishes since $\tau $ is a
constant -- the extrinsic curvature no longer provides a measure of time
evolution.

The solution of (\ref{N0p1})\ is

\begin{equation}
\Psi =-2\;\ln \left\{ Ax\exp (t/\ell )/\ell -\beta (t)\right\} \ \ \ \text{%
and \ }\ \Pi =\frac{\exp \left( t/\ell \right) }{\kappa \ell \left( Ax\exp
(t/\ell )/\ell -\beta (t)\right) }\text{ \ \ \ \ \ .}  \label{psisoltrump}
\end{equation}%
It is not possible to make $\Psi $ periodic unless $A=0$ . The remaining
equations then force the solution 
\begin{equation}
\Psi =-2t/\ell \ \ \ \text{and \ }\ \Pi =0\text{ \ \ \ \ \ }
\label{psisoltrump2}
\end{equation}%
and the period is arbitrary.

\subsubsection{Comment on the $N_{0}=1$ de Sitter solutions}

The 3 solutions (\ref{dsmet1},\ref{dsmet2},\ref{dsmet3}) represent de Sitter
spacetime in different coordinates. \ The maximally extended solution is (%
\ref{dsmet1})\ . All solutions are locally transformable into each other and
are all physicallly equivalent \ when the spatial direction is not compact (%
{\it ie }$x$ is not periodic). Given the metrics

\begin{eqnarray}
&&  \nonumber \\
ds_{\text{candlestick}}^{2} &=&-dt^{2}+\cosh ^{2}\left( t/\ell \right)
\,dx^{2}  \label{desit1a} \\
ds_{\text{dish}}^{2} &=&-dT^{2}+\sinh ^{2}\left( T/\ell \right) \,dX^{2}
\label{desit1b} \\
ds_{\text{trumpet}}^{2} &=&-d{\cal T}^{2}+\exp \left( 2{\cal T}/\ell \right)
\,d{\cal X}^{2}  \label{desit1c}
\end{eqnarray}%
the transformations are 
\begin{eqnarray}
&&\sinh \left( t/\ell \right) =\sinh \left( T/\ell \right) \cosh \left(
X/\ell \right) =\left( 1+\frac{{\cal X}^{2}}{2\ell ^{2}}\right) \sinh ({\cal %
T}/l)+\frac{{\cal X}^{2}}{2\ell ^{2}}\cosh ({\cal T}/l)  \label{desittfmV} \\
&&\cosh \left( t/\ell \right) \cos (x/\ell )=\cosh \left( T/\ell \right) =%
\frac{{\cal X}}{\ell }\exp ({\cal T}/l)  \label{desittfmX} \\
&&\cosh \left( t/\ell \right) \sin (x/\ell )=\sinh \left( T/\ell \right)
\sinh \left( X/\ell \right) =\left( 1-\frac{{\cal X}^{2}}{2\ell ^{2}}\right)
\cosh ({\cal T}/l)-\frac{{\cal X}^{2}}{2\ell ^{2}}\sinh ({\cal T}/l)
\label{desittfmW}
\end{eqnarray}

However when $x$ {\it is }periodic the solutions are physically distinct. \
The dish and trumpet solutions have arbitrary periodicity whereas the
candlestick solution can have only discrete periodicity. \ Note that these
constraints are due to the behaviour of the $\Psi $ field, and are not
dictated by the metric. The candlestick and dish solutions have a
singularity in $\Psi $\ at $t=0$ .

Hence even though the gravity/matter system (\ref{eq-z}) and (\ref{RT}) is
closed, the global properties of the dilaton field can influence the
spacetime through self-consistency of the solutions to (\ref{eq-Psi}).

\bigskip

\subsection{The $c^{2}$ Solutions}

\bigskip Next I shall solve equations (\ref{N0p1} --\ref{n0p5}) in terms of
the quantity 
\begin{equation}
c^{2}=\gamma \left( \tau ^{2}-\frac{\Lambda }{2}\right)  \label{csq}
\end{equation}%
which can be positive ($c_{+}^{2}=c^{2}>0$), zero ($c=0$) or negative ($%
c_{-}^{2}=-c^{2}>0$). \ Although this approach will not yield any new
solutions, it can be straightforwardly extended to the single-particle case,
and so is instructive to consider.

\subsubsection{$c_{+}^{2}=c^{2}>0$}

The solution of (\ref{N0p2} ) is 
\begin{equation}
N_{0}=\hat{N}\cosh \left( c_{+}x+\vartheta \right) -\frac{\gamma \dot{\tau}}{%
c_{+}^{2}}=-\frac{\gamma \dot{\tau}}{c_{+}^{2}}  \label{N0-0pcp}
\end{equation}%
where periodicity in $x$ forces $\hat{N}=0$. The solution of (\ref{N0p1} )
is then 
\begin{equation}
\Psi =-2\ln \left\{ \beta -\cosh (c_{+}x+D)\right\} +\ln \left( \frac{%
c_{+}^{2}(\beta ^{2}-1)}{(\kappa \Pi _{0})^{2}}\right)  \label{psi-0pcp}
\end{equation}%
where $D(t)$ and $\beta (t)$ are constants of integration. \ Periodicity in $%
\Psi $ forces $D=0$ and periodicity in\ $\Psi ^{\prime }$forces $\beta
\rightarrow \infty $ so that\bigskip 
\begin{equation}
\Psi =\ln \left( \frac{c_{+}^{2}}{(\kappa \Pi _{0})^{2}}\right)
\label{psi-0p2cp}
\end{equation}%
with the result that $\Psi $ is independent of $x$. \ 

Periodicity in $x$ forces $N_{1}=$constant (which can be chosen to be $0$)
from (\ref{n0p3}), which in turn yields $c_{+}^{2}=c_{+0}^{2}=$ constant,
which is 
\begin{equation}
\sqrt{\gamma }=\frac{c_{+0}}{\sqrt{\tau ^{2}-\frac{\Lambda }{2}}}
\label{gam-0psol}
\end{equation}%
Then (\ref{n0p4}), (\ref{psi-0p2cp}) and (\ref{n0p5}) yield $\Pi =\pm
c_{+0}/\kappa $ which gives one of two possible solutions for $\Psi $%
\begin{equation}
\Psi =\left\{ 
\begin{array}{c}
2\ln \left( L^{2}\tau \sqrt{\tau ^{2}-\frac{\Lambda }{2}}+L^{2}\left( \tau
^{2}-\frac{\Lambda }{2}\right) \right) \text{ \ \ \ \ }\Pi =+c_{+0}/\kappa
\\ 
-2\ln \left( \frac{\tau }{\sqrt{\tau ^{2}-\frac{\Lambda }{2}}}+1\right) 
\text{ \ \ \ \ \ \ \ \ \ \ \ \ \ \ \ \ \ \ \ \ \ \ \ \ \ \ \ \ \ }\Pi
=-c_{+0}/\kappa%
\end{array}%
\right. \   \label{gam-0psisol}
\end{equation}%
where the arbitrary constant of integration has been chosen to be $L$. \ 

To summarize, the solution is 
\begin{eqnarray}
&&  \nonumber \\
ds^{2} &=&-\frac{\dot{\tau}^{2}dt^{2}}{\left( \tau ^{2}-\frac{\Lambda }{2}%
\right) ^{2}}+\frac{c_{+0}^{2}dx^{2}}{\left( \tau ^{2}-\frac{\Lambda }{2}%
\right) }  \label{cplus1} \\
\Psi &=&2\ln \left( L^{2}\left( \tau ^{2}-\frac{\Lambda }{2}\right) \pm
L^{2}\tau \sqrt{\tau ^{2}-\frac{\Lambda }{2}}\right) \text{ \ \ \ \ \ }\Pi
=\pm c_{+}/\kappa \text{ }  \label{cplus2}
\end{eqnarray}%
Setting $\Lambda =-2/\ell ^{2}$ and $\tau =\tan \left( t/\ell \right) /\ell $
yields from (\ref{cplus1},\ref{cplus2}) the solution (\ref{adsmet},\ref%
{adspsi}), \ whereas setting $\Lambda =0$ and $\tau =-1/t$\ gives the
solution (\ref{flatcirmet},\ref{psiflatsol}) (provided $\Psi $ is rescaled
by a divergent constant). The dish solution (\ref{dsmet2},\ref{psidish}) is
recoved by setting $\Lambda =2/\ell ^{2}$ and $\tau =-\coth \left( t/\ell
\right) /\ell $ . The Hamiltonian is 
\begin{equation}
H=\frac{4Lc_{+0}\dot{\tau}}{\kappa \sqrt{\tau ^{2}-\frac{\Lambda }{2}}}=%
\frac{4c_{+0}L}{\kappa \ell \sqrt{\left( t/\ell \right) ^{2}-\epsilon }}
\label{cplusH}
\end{equation}%
where the latter equality follows if $\tau =t/\ell ^{2}$, \ and $\epsilon =$%
sgn$\left( \Lambda \right) $, with $\epsilon $ vanishing if $\Lambda =0$. \ 

If $\ \tau $ is chosen so that$\ \ell \dot{\tau}=\sqrt{\tau ^{2}-\frac{%
\Lambda }{2}}$ , then the Hamiltonian 
\begin{equation}
H=\frac{4L}{\kappa \ell }c_{+0}  \label{cplusH2}
\end{equation}
and is constant, and the metric is%
\begin{equation}
ds^{2}=\frac{-dt^{2}+c_{+0}^{2}dx^{2}}{\left( \tau ^{2}-\frac{\Lambda }{2}%
\right) }  \label{cplus3}
\end{equation}%
which is conformal to a flat metric of topology ${\Bbb R}\times S^{1}$.

\subsubsection{$c^{2}=0$}

In this case $\tau =1/\ell =\sqrt{2/\Lambda }$, and so the solution is a bit
different. \ Solving for $\gamma $ from (\ref{N1p1}) gives 
\begin{equation}
\dot{\gamma}/2-\frac{\gamma }{\ell }=0\Rightarrow \gamma =\exp \left( t/\ell
\right)  \label{c01}
\end{equation}%
The constraint equation for $\Psi $ is 
\begin{eqnarray*}
\Psi ^{\prime \prime }-\frac{1}{4}(\Psi ^{\prime })^{2}-(\kappa \Pi
_{0}(t))^{2}\exp [\Psi ] &=&0 \\
\Psi =-2\ln \phi \Rightarrow \phi \frac{d}{d\phi }\left( \left( \phi
^{\prime }\right) ^{2}\right) -\left( \phi ^{\prime }\right) ^{2}+\left(
\kappa \Pi _{0}\right) ^{2} &=&0
\end{eqnarray*}%
which has the solution 
\begin{eqnarray}
\phi &=&\left[ \frac{\sigma }{\alpha (t)}\left( \frac{\alpha (t)}{2}x+\beta
(t)\right) ^{2}-\left( \kappa \Pi _{0}\right) ^{2}\right]  \nonumber \\
&\Rightarrow &\Psi =-2\ln \left( \left[ \frac{\sigma }{\alpha (t)}\left( 
\frac{\alpha (t)}{2}x+\beta (t)\right) ^{2}-\left( \kappa \Pi _{0}\right)
^{2}\right] \right)  \label{c02}
\end{eqnarray}%
This can't be periodic unless $\alpha =0$, which in turn forces $\Pi _{0}=0$%
. \ The other equations then finally yield the solution%
\begin{eqnarray}
ds^{2} &=&-dt^{2}+\exp \left( 2t/\ell \right) \,dx^{2}\text{ \ \ }
\label{c03} \\
\Psi &=&-2t/\ell \ \ \ \text{and \ }\ \Pi =0\text{ .}  \label{c04}
\end{eqnarray}%
which is the trumpet solution (\ref{desit1c}). The Hamiltonian vanishes as
discussed above.

\subsubsection{$c_{-}^{2}=-c^{2}>0$}

In this case $c_{-}^{2}=\gamma \left( \frac{\Lambda }{2}-\tau ^{2}\right) $%
\bigskip $=-c^{2}$. The solution of (\ref{N0p2} ) is now 
\begin{equation}
N_{0}=\hat{N}(t)\cos \left( c_{-}x+\vartheta \right) +\frac{\gamma \dot{\tau}%
}{c_{-}^{2}}  \label{N0-0pcn}
\end{equation}%
where periodicity in $x$ no longer forces $\hat{N}=0$. The solution of (\ref%
{N0p1} ) is then 
\begin{equation}
\Psi =-2\ln \left\{ \cos (c_{-}x+D)+\beta \right\} +\ln \left( \frac{%
c_{-}^{2}(1-\beta ^{2})}{(\kappa \Pi _{0})^{2}}\right)  \label{psicmintemp}
\end{equation}%
where $D(t)$ and $\beta (t)$ are constants of integration. \ The solution to
(\ref{n0p3}) is 
\[
N_{1}=(\dot{\gamma}/2-\gamma \tau \frac{\gamma \dot{\tau}}{c_{-}^{2}})x-%
\frac{\gamma \tau }{c_{-}}\hat{N}\sin \left( c_{-}x+\vartheta \right) =\frac{%
\gamma \dot{c}_{-}}{c_{-}}x-\frac{\gamma \tau }{c_{-}}\hat{N}\sin \left(
c_{-}x+\vartheta \right) 
\]

Periodicity and smoothness in $x$ implies $N_{0}^{\prime
}(L,t)=N_{0}^{\prime }(-L,t)=0$ $,$ $N_{0}(L)=N_{0}(-L)=0$ from (\ref{n0p3}%
), yielding $c_{-}L=N\pi $, or%
\begin{equation}
\sqrt{\gamma }=\frac{N\pi }{L\sqrt{\left( \frac{\Lambda }{2}-\tau
^{2}\right) }}  \label{cminus0p}
\end{equation}%
where $N$ is a positive integer. \ Periodicity in $N_{1}\left( x\right) $
then forces $\vartheta $ to be an integer multiple of $\pi $ as well, and
without loss of generality one can take $\vartheta =0$, giving

\begin{equation}
N_{1}=-\frac{L\gamma \tau }{N\pi }\hat{N}(t)\sin \left( N\pi \frac{x}{L}%
\right)  \label{N1-0pcm}
\end{equation}

Solving the remaining equations is made easier by setting $\ l=L/N\pi $ , $%
D=0$ and choosing $\hat{N}(t)=l^{2}\gamma \dot{\tau}/\beta \left( t\right) .$
This gives%
\begin{eqnarray}
0 &=&\dot{\Pi}+\partial _{1}(-\frac{1}{\gamma }N_{1}\Pi +\frac{1}{2\kappa 
\sqrt{\gamma }}N_{0}\Psi ^{\prime }+\frac{1}{\kappa \sqrt{\gamma }}%
N_{0}^{\prime })  \nonumber \\
&=&\dot{\Pi}+\partial _{1}(-\frac{1}{\gamma }N_{1}\Pi )  \nonumber \\
&=&\partial _{0}\left( \frac{\sqrt{(1-\beta ^{2})}}{(l\left( \cos
(x/l)+\beta \right) )}\right) +\frac{\gamma \dot{\tau}\tau l^{2}}{\beta }%
\left[ \frac{\sin \left( x/l\right) \sqrt{(1-\beta ^{2})}}{(\cos (x/l)+\beta
)}\right] ^{\prime }  \nonumber \\
&=&-\frac{\beta \dot{\beta}}{l\left( \cos (x/l)+\beta \right) \sqrt{(1-\beta
^{2})}}-\frac{\dot{\beta}\sqrt{(1-\beta ^{2})}}{l\left( \cos (x/l)+\beta
\right) ^{2}}  \nonumber \\
&&+\frac{\gamma \dot{\tau}\tau l}{\beta }\sqrt{(1-\beta ^{2})}\left[ \frac{%
\cos \left( x/l\right) }{(\cos (x/l)+\beta )}+\frac{\sin ^{2}\left(
x/l\right) }{(\cos (x/l)+\beta )^{2}}\right]  \nonumber \\
&=&\frac{1}{l\left( \cos (x/l)+\beta \right) \sqrt{(1-\beta ^{2})}}\left[
-\beta \dot{\beta}+\gamma \dot{\tau}\tau (1-\beta ^{2})l^{2}\right] \left( 
\frac{1}{\beta }+\cos (x/l)\right)  \label{cnegPi}
\end{eqnarray}%
which using (\ref{cminus0p}) implies 
\begin{equation}
\beta =\sqrt{1-{\frak l}^{2}\left( \frac{\Lambda }{2}-\tau ^{2}\right) }
\label{cnegbeta}
\end{equation}%
where ${\frak l}$ is an arbitrary constant. \ The remaining equation is 
\begin{eqnarray}
0 &=&\dot{\Psi}+2N_{0}\left( \kappa \frac{\Pi }{\sqrt{\gamma }}+\tau \right)
-N_{1}(\frac{1}{\gamma }\Psi ^{\prime })  \nonumber \\
&=&\frac{-2\dot{\beta}}{\left( \cos (x/l)+\beta \right) }-\frac{2\beta \dot{%
\beta}}{(1-\beta ^{2})}-2\frac{\dot{\Pi}_{0}}{\Pi _{0}}  \nonumber \\
&&+2\frac{l^{2}\gamma \dot{\tau}}{\beta }\left( \cos \left( x/l\right)
+\beta \right) \left[ \frac{\pm \sqrt{(1-\beta ^{2})}}{\sqrt{\gamma }%
(l\left( \cos (x/l)+\beta \right) )}+\tau \right] +\frac{2\gamma \dot{\tau}%
\tau l^{2}}{\beta }\frac{\sin ^{2}\left( x/l\right) }{\left( \cos
(x/l)-\beta \right) }  \nonumber \\
&=&\frac{2}{\left( \cos (x/l)+\beta \right) }\left[ -\dot{\beta}+\frac{%
\gamma \dot{\tau}\tau l^{2}}{\beta }\sin ^{2}\left( x/l\right) +\frac{%
l^{2}\gamma \dot{\tau}}{\beta }\left( \cos \left( x/l\right) +\beta \right)
\left( \frac{\pm \sqrt{(1-\beta ^{2})}}{\sqrt{\gamma }l}+\tau \left( \cos
(x/l)+\beta \right) \right) \right]  \nonumber \\
&&-\frac{2\beta \dot{\beta}}{(1-\beta ^{2})}-2\frac{\dot{\Pi}_{0}}{\Pi _{0}}
\nonumber \\
&=&\frac{2}{\left( \cos (x/l)+\beta \right) }\left[ \left( \cos (x/l)+\beta
\right) \left( +2\gamma \dot{\tau}\tau l^{2}\pm \frac{l\sqrt{\gamma }\dot{%
\tau}\sqrt{(1-\beta ^{2})}}{\beta }\right) \right] -2\gamma \dot{\tau}\tau
l^{2}-2\frac{\dot{\Pi}_{0}}{\Pi _{0}}  \nonumber \\
&=&2\frac{\dot{\tau}\tau }{\frac{\Lambda }{2}-\tau ^{2}}\pm 2\frac{{\frak l}%
\dot{\tau}}{\sqrt{1-{\frak l}^{2}\left( \frac{\Lambda }{2}-\tau ^{2}\right) }%
}-2\frac{\dot{\Pi}_{0}}{\Pi _{0}}  \label{cnegpsidot}
\end{eqnarray}%
which is the equation that determines $\Pi _{0}$. \ It gives 
\begin{equation}
\Pi _{0}=\frac{\left( {\frak l}\tau +\sqrt{1-{\frak l}^{2}\left( \frac{%
\Lambda }{2}-\tau ^{2}\right) }\right) ^{\pm 1}}{\sqrt{\frac{\Lambda }{2}%
-\tau ^{2}}}  \label{cnegeqs}
\end{equation}

\bigskip

Hence the solution takes the form 
\begin{eqnarray}
&&  \nonumber \\
N_{0} &=&\frac{\dot{\tau}}{\left( \frac{\Lambda }{2}-\tau ^{2}\right) \sqrt{%
1-{\frak l}^{2}\left( \frac{\Lambda }{2}-\tau ^{2}\right) }}\left( \cos
\left( \frac{x}{l}\right) +\sqrt{1-{\frak l}^{2}\left( \frac{\Lambda }{2}%
-\tau ^{2}\right) }\right)  \label{cnegN0} \\
\text{\ \ \ }N_{1} &=&\frac{\tau \dot{\tau}}{l\left( \frac{\Lambda }{2}-\tau
^{2}\right) ^{2}\sqrt{1-{\frak l}^{2}\left( \frac{\Lambda }{2}-\tau
^{2}\right) }}\sin \left( \frac{x}{l}\right)  \label{cnegN1} \\
\Psi &=&-2\ln \left\{ \cos \left( x/l\right) +\sqrt{1-{\frak l}^{2}\left( 
\frac{\Lambda }{2}-\tau ^{2}\right) }\right\} +2\ln \left( \frac{{\frak l}%
\left( {\frak l}\tau \mp \sqrt{1-{\frak l}^{2}\left( \frac{\Lambda }{2}-\tau
^{2}\right) }\right) }{\kappa l}\right)  \label{cnegPsi} \\
\Pi &=&\frac{\pm {\frak l}\sqrt{\frac{\Lambda }{2}-\tau ^{2}}}{\kappa \left(
\cos \left( \frac{x}{l}\right) +\sqrt{1-{\frak l}^{2}\left( \frac{\Lambda }{2%
}-\tau ^{2}\right) }\right) }  \label{cnegPisol}
\end{eqnarray}%
where $L=N\pi l$\ , and the spatial metric $\gamma $ \ is given by (\ref%
{cminus0p}). It is equivalent to the candlestick solution (\ref{dsmet1}) in
different coordinates. In these coordinates the metric and lapse function
diverge at $\tau =\pm \sqrt{\frac{\Lambda }{2}}$.

The Hamiltonian is equal to%
\begin{equation}
H=\frac{4\dot{\tau}}{\kappa }\frac{N\pi }{\sqrt{\left( \frac{\Lambda }{2}%
-\tau ^{2}\right) }}  \label{Hamcandle}
\end{equation}%
and is the same as that given in (\ref{candleH}) provided $\ \tau =\tanh
\left( t/\ell \right) /\ell $\ and \ $L=N\pi \ell $. \ Alternatively for $%
\tau =\cos \left( t/\ell \right) /\ell $, the Hamiltonian is constant. \ The
topology is the same as in the candlestick case: there is a countably
infinite set of solutions, each labelled by $N$, in which the circle
contracts from large radius to some minimal size and then expands out again
to infinity. \ 

\bigskip

\section{The 1-particle case}

When there is 1 particle, set $p=0$\thinspace , $m=M$ and $z=0$. Equation (%
\ref{R1}) is again trivially solved to yield 
\begin{equation}
\Pi =\Pi _{0}(t)\exp [\Psi /2]  \label{Pisol1p}
\end{equation}%
and the system must be solved for the cases ($c^{2}=c_{+}^{2}>0$), zero ($%
c=0 $) or negative ($-c^{2}=c_{-}^{2}>0$).

\bigskip

\subsection{Derivation of Single-particle solutions}

\subsubsection{$c_{+}^{2}=c^{2}>0$}

\bigskip

Equations (\ref{fin1a}) to (\ref{fin8})\ \ now reduce to 
\begin{eqnarray}
&&  \nonumber \\
&&\Psi ^{\prime \prime }-\frac{1}{4}(\Psi ^{\prime })^{2}-(\kappa \Pi
_{0}(t))^{2}\exp [\Psi ]+\gamma \left( \tau ^{2}-\Lambda /2\right) +\kappa 
\sqrt{\gamma }M\delta (x)=0  \label{N1p1} \\
&&N_{0}^{\prime \prime }=\overset{\cdot }{\tau }\gamma +N_{0}\left\{ (\tau
^{2}-\Lambda /2)\gamma +\kappa \sqrt{\gamma }\frac{M}{2}\;\delta (x)\right\}
\label{N1p2} \\
&&N_{1}^{\prime }=\dot{\gamma}/2-\gamma \tau N_{0}  \label{N1p3} \\
&&\dot{\Pi}+\partial _{1}(-\frac{1}{\gamma }N_{1}\Pi +\frac{1}{2\kappa \sqrt{%
\gamma }}N_{0}\Psi ^{\prime }+\frac{1}{\kappa \sqrt{\gamma }}N_{0}^{\prime
})=0  \label{N1p4} \\
&&\dot{\Psi}+2N_{0}\left( \kappa \frac{\Pi }{\sqrt{\gamma }}+\tau \right)
-N_{1}(\frac{1}{\gamma }\Psi ^{\prime })=0  \label{N1p5psi} \\
&&\left. \frac{\partial N_{0}}{\partial x}\right| _{x=0}=0\text{ \ \ \ \ \ \
\ \ \ \ \ \ \ \ }N_{1}(0)=0  \label{N1p6}
\end{eqnarray}%
To solve these equations, note that (\ref{N1p2}) becomes 
\begin{equation}
\triangle N_{0}=\gamma \dot{\tau}+c_{+}^{2}N_{0}+\frac{1}{2}\kappa M\sqrt{%
\gamma }\delta (x)N_{0}\;  \label{N01pdbleprime}
\end{equation}%
and has the solution 
\begin{equation}
N_{0}(x,t)=\frac{\gamma \dot{\tau}}{c_{+}^{2}\beta (t)}\left\{ \cosh
(c_{+}(|x|-L))-\beta (t)\right\} \;\;  \label{N01psol}
\end{equation}%
where 
\begin{equation}
\beta =\cosh (c_{+}L)+\frac{4c_{+}}{\kappa M\sqrt{\gamma }}\sinh (c_{+}L)%
\text{ \ \ .}  \label{beta1peq}
\end{equation}%
Note that $N_{0}(L,t)=N_{0}(-L,t)$ and $N_{0}^{\prime }(L,t)=N_{0}^{\prime
}(-L,t)=0$ , as respectively required by periodicity and smoothness.
Furthermore $N_{0}^{\prime }(0,t)=0$ as required by (\ref{N1p6}). \ \
Solving (\ref{N1p1}) yields

\begin{eqnarray}
&&  \nonumber \\
\Psi &=&-2\ln \left\{ \cosh (c_{+}(|x|-L))-\beta (t)\right\} +\Psi _{0}
\label{psi1psol} \\
\Psi _{0} &=&-2\ln \left( \frac{\kappa \Pi _{0}(t)}{c_{+}\sqrt{\beta ^{2}-1}}%
\right) \text{ \ \ \ \ \ \ \ \ \ \ \ \ \ \ \ \ \ \ \ \ \ \ .}
\label{psi01psol}
\end{eqnarray}%
which also has the requisite periodicity. Similarly, using \ (\ref{Pisol1p}%
), 
\begin{equation}
\kappa \Pi =\frac{\pm c_{+}\sqrt{\beta ^{2}-1}}{\cosh (c_{+}(|x|-L))-\beta
(t)}  \label{Pisoldbl1p}
\end{equation}

\bigskip

Using (\ref{N01psol}) the solution to (\ref{N1p3}) for $N_{1}$ is 
\begin{eqnarray}
N_{1}(x,t) &=&\frac{1}{2}\dot{\gamma}x-\gamma \tau \frac{\gamma \dot{\tau}}{%
c_{+}^{3}\beta }\left\{ {\mbox{sgn}}(x)\left[ \sinh (c_{+}(|x|-L))+\sinh
(c_{+}L)\right] -\beta c_{+}x\right\}  \nonumber \\
&=&\gamma \frac{\dot{c}_{+}}{c_{+}}x-\gamma \tau \frac{\gamma \dot{\tau}}{%
c_{+}^{3}\beta }{\mbox{sgn}}(x)\left[ \sinh (c_{+}(|x|-L))+\sinh (c_{+}L)%
\right]  \label{N11psol}
\end{eqnarray}%
which satisfies (\ref{N1p6}).

\bigskip

For $\Psi $, $\Pi $ and $N_{0}$ the periodicity conditions (\ref{smooth})
are satisfied. However although $N_{1}^{\prime }(L)=N_{1}^{\prime }(-L)=0$,
it is also necessary to impose $N_{1}(L)=N_{1}(-L)=0$. This requirement
allows $\gamma (t)$ to be determined by the equation 
\begin{equation}
\beta c_{+}^{2}\dot{c}_{+}L-\gamma (\tau \dot{\tau})\sinh (c_{+}L)=0
\label{betadoteq}
\end{equation}%
which, upon using the definitions of $c_{+}$ and $\beta $ yields after
integration%
\begin{equation}
\cosh (c_{+}L)+\frac{\kappa M\sqrt{\gamma }}{4c_{+}}\sinh (c_{+}L)=\cosh (L%
\sqrt{\gamma }\sqrt{\tau ^{2}-\frac{\Lambda }{2}})+\frac{\kappa M}{4\sqrt{%
\tau ^{2}-\frac{\Lambda }{2}}}\sinh (L\sqrt{\gamma }\sqrt{\tau ^{2}-\frac{%
\Lambda }{2}})=\xi  \label{cons1p}
\end{equation}%
where $\xi $ is a constant and eq. (\ref{csq}) has been employed.\ 

Eq. (\ref{cons1p}) can be solved for $\sqrt{\gamma }$, yielding $\gamma $ as
a function of $\tau $ and the parameters $M$ and $\xi $. \ The solution is
given by finding the intersection of the curve $\cosh (\chi )+m(\tau )\sinh
(\chi )$, (where $\chi (\tau )=L\sqrt{\gamma }\sqrt{\tau ^{2}-\frac{\Lambda 
}{2}}>0$ and $m(\tau )=\frac{\kappa M}{4\sqrt{\tau ^{2}-\frac{\Lambda }{2}}}$%
) with a horizontal line of height $\xi $. For large $\tau $, the term
proportional to $M$ becomes negligible, and only for $\xi >1$ do admissible
solutions exist. \ As $\tau $ decreases from infinity, $m(\tau )$ increases
from zero, the curve becomes steeper for $\chi >0$ and so the point of
intersection $\chi _{0}$ decreases. \ 

The solution to (\ref{cons1p}) is%
\begin{equation}
\sqrt{\gamma }=\frac{4}{\kappa ML}m(\tau )\text{arctanh}\left( \frac{\xi 
\sqrt{m^{2}(\tau )+(\xi ^{2}-1)}-m(\tau )}{m^{2}(\tau )+\xi ^{2}}\right)
\label{gam1pmtau}
\end{equation}%
or, more explicitly, 
\begin{equation}
\sqrt{\gamma }=\frac{1}{L\sqrt{\tau ^{2}-\Lambda /2}}\text{arctanh}\left( 
\frac{\xi \sqrt{\frac{\kappa ^{2}M^{2}}{16}+(\xi ^{2}-1)(\tau ^{2}-\Lambda
/2)}-\frac{\kappa M}{4}}{\frac{\kappa ^{2}M^{2}}{16}+\xi ^{2}(\tau
^{2}-\Lambda /2)}\sqrt{\tau ^{2}-\Lambda /2}\right)  \label{gam1psol}
\end{equation}%
which reduces to the expression (\ref{gam-0psol}) as $M\rightarrow 0$. \ The
Hamiltonian now takes on the general form 
\begin{equation}
H=\frac{4\dot{\tau}}{\kappa \sqrt{\tau ^{2}-\frac{\Lambda }{2}}}\text{arctanh%
}\left( \frac{\xi \sqrt{\frac{\kappa ^{2}M^{2}}{16}+(\xi ^{2}-1)(\tau
^{2}-\Lambda /2)}-\frac{\kappa M}{4}}{\frac{\kappa ^{2}M^{2}}{16}+\xi
^{2}(\tau ^{2}-\Lambda /2)}\sqrt{\tau ^{2}-\Lambda /2}\right)
\label{cplusHt1p}
\end{equation}%
and the function $\beta (\tau )$ is also found to be 
\begin{equation}
\beta =\frac{4}{\kappa M}\sqrt{\frac{\kappa ^{2}M^{2}}{16}+(\xi ^{2}-1)(\tau
^{2}-\Lambda /2)}  \label{beta1psol}
\end{equation}

\bigskip

\bigskip The remaining task is to solve equations (\ref{N1p4}) and (\ref%
{N1p5psi}). \ The first of these simplifies to 
\begin{equation}
\frac{\partial }{\partial t}\ln \left( \kappa \Pi \right) -\left[ \frac{1}{%
\gamma }N_{1}\left( \ln \Pi \right) \right] ^{\prime }=0  \label{simpleN1p4}
\end{equation}%
since $N_{0}\exp \left( \Psi /2\right) $ is independent of $x$. \ Inserting (%
\ref{Pisoldbl1p}) and (\ref{N11psol}) into (\ref{simpleN1p4}) yields 
\begin{eqnarray*}
&&\frac{\partial }{\partial t}\ln \left( \kappa \Pi \right) -\frac{%
N_{1}^{\prime }}{\gamma }-\frac{N_{1}}{\gamma }\left[ \ln \left( \kappa \Pi
\right) \right] ^{\prime } \\
&=&\left( \frac{\dot{c}_{+}}{c_{+}}-\frac{\dot{\gamma}}{2\gamma }-\frac{%
\gamma \dot{\tau}\tau }{c_{+}^{2}}\right) +\frac{1}{\left( \cosh
(c_{+}(|x|-L))-\beta \right) }\left[ \sinh (c_{+}(|x|-L))\left( c_{+}\dot{c}%
L-\sinh (c_{+}L)\frac{\gamma \dot{\tau}\tau }{c_{+}\beta }\right) \right. \\
&&\left. +\frac{\gamma \dot{\tau}\tau }{c_{+}^{2}\beta }\left( \cosh
(c_{+}(|x|-L))\left( \cosh (c_{+}(|x|-L))-\beta \right) -\sinh
^{2}(c_{+}(|x|-L))+\dot{\beta}\right) \right] +\frac{\beta \dot{\beta}}{%
\beta ^{2}-1} \\
&=&\frac{\left( \beta \cosh (c_{+}(|x|-L)-1\right) }{\beta \left( \cosh
(c_{+}(|x|-L))-\beta \right) }\left[ \left( \frac{\beta \dot{\beta}}{\beta
^{2}-1}-\frac{\gamma \dot{\tau}\tau }{c_{+}^{2}}\right) \right] \\
&=&0
\end{eqnarray*}%
where \ (\ref{csq}) and (\ref{betadoteq}) were used to simplify the second
line, and the last line follows from (\ref{beta1psol}). \ Finally, inserting
(\ref{Pisol1p}) into (\ref{N1p5psi}) and using (\ref{N1p4}) yields%
\begin{equation}
\dot{\Pi}_{0}-\Pi _{0}\left[ \frac{\kappa \Pi _{0}}{\sqrt{\gamma }}\left(
N_{0}\exp [\Psi /2]\right) +\frac{\dot{\gamma}}{2\gamma }\right] +\frac{\exp
[-\Psi /2]}{\kappa \sqrt{\gamma }}\partial _{1}(\exp [-\Psi /2]\left(
N_{0}\exp [\Psi /2]\right) ^{\prime })=0  \label{newPidot}
\end{equation}%
which in turn simplifies to 
\begin{equation}
\frac{\partial }{\partial t}\ln \left( \frac{\kappa \Pi _{0}}{\sqrt{\gamma }}%
\right) \mp \frac{\sqrt{\gamma }\dot{\tau}}{\beta c_{+}}\sqrt{\beta ^{2}-1}=%
\frac{\partial }{\partial t}\ln \left( \frac{\kappa \Pi _{0}}{\sqrt{\gamma }}%
\right) \mp \frac{\dot{\tau}\sqrt{(\xi ^{2}-1)}}{\sqrt{\frac{\kappa ^{2}M^{2}%
}{16}+(\xi ^{2}-1)(\tau ^{2}-\Lambda /2)}}=0  \label{finPidot}
\end{equation}%
upon insertion of the solutions (\ref{N01psol},\ref{psi1psol}) into (\ref%
{newPidot}). \ This determines $\Pi _{0}$ as a function of $t$, \ and
integrates to 
\begin{equation}
\kappa \Pi _{0}\left( \tau \right) =\sqrt{\gamma }{\frak L}^{2}\left( \tau
\pm \sqrt{\frac{\kappa ^{2}M^{2}}{16(\xi ^{2}-1)}+(\tau ^{2}-\Lambda /2)}%
\right)  \label{finPi0}
\end{equation}%
where the dimensionless constant of integration has been chosen to be $%
{\frak L}^{2}$.

All equations are satisfied, and so the single particle solution for $%
c^{2}>0 $ is 
\begin{eqnarray}
&&  \nonumber \\
N_{0}(x,t) &=&\frac{\dot{\tau}}{(\tau ^{2}-\Lambda /2)}\left\{ \frac{\kappa
M\cosh (\tanh ^{-1}\left( \sigma \left( \tau \right) \right) (|x|/L-1))}{4%
\sqrt{\frac{\kappa ^{2}M^{2}}{16}+(\xi ^{2}-1)(\tau ^{2}-\Lambda /2)}}%
-1\right\}  \label{1psola} \\
N_{1}(x,t) &=&\frac{\kappa M{\mbox{sgn}}(x)\tanh ^{-1}\left( \sigma \left(
\tau \right) \right) \tau \dot{\tau}\left[ \sinh (\tanh ^{-1}\left( \sigma
\left( \tau \right) \right) (|x|/L-1))+\frac{\sigma \left( \tau \right)
(|x|/L-1)}{\sqrt{1-\sigma ^{2}\left( \tau \right) }}\right] }{4L(\tau
^{2}-\Lambda /2)^{2}\sqrt{\frac{\kappa ^{2}M^{2}}{16}+(\xi ^{2}-1)(\tau
^{2}-\Lambda /2)}}  \label{1psolb} \\
\Psi (x,t) &=&-2\ln \left\{ \frac{\kappa M\cosh (\tanh ^{-1}\left( \sigma
\left( \tau \right) \right) (|x|/L-1))}{4\sqrt{\frac{\kappa ^{2}M^{2}}{16}%
+(\xi ^{2}-1)(\tau ^{2}-\Lambda /2)}}-1\right\}  \nonumber \\
&&-2\ln \left( \frac{{\frak L}^{2}\left[ \tau \pm \sqrt{\frac{\kappa
^{2}M^{2}}{16(\xi ^{2}-1)}+(\tau ^{2}-\Lambda /2)}\right] }{(\tau
^{2}-\Lambda /2)}\sqrt{\frac{\kappa ^{2}M^{2}}{16(\xi ^{2}-1)}+(\tau
^{2}-\Lambda /2)}\right)  \label{1psolc} \\
\kappa \Pi &=&\frac{\pm \sqrt{\xi ^{2}-1}\sqrt{(\tau ^{2}-\Lambda /2)}\tanh
^{-1}\left( \sigma \left( \tau \right) \right) }{\frac{\kappa M}{4}\cosh
(c_{+}(|x|-L))-\sqrt{\frac{\kappa ^{2}M^{2}}{16}+(\xi ^{2}-1)(\tau
^{2}-\Lambda /2)}}  \label{1psolPi}
\end{eqnarray}%
where $\sigma \left( \tau \right) =\frac{\xi \sqrt{\frac{\left( \kappa M\ell
\right) ^{2}}{16}+(\xi ^{2}-1)(\tau ^{2}-\Lambda /2)}-\frac{\kappa M\ell }{4}%
}{\frac{\left( \kappa M\ell \right) ^{2}}{16}+\xi ^{2}(\tau ^{2}-\Lambda /2)}%
\sqrt{\tau ^{2}-\Lambda /2}$ and the spatial metric $\gamma $ is given in (%
\ref{gam1psol}).

\subsubsection{$c_{+}^{2}=0$}

In this case $\tau =\sqrt{\Lambda /2}$, which in turn implies 
\begin{equation}
N_{0}^{\prime \prime }=\kappa \sqrt{\gamma }\frac{M}{2}\;N_{0}\;\delta (x)
\label{N0cp0eq}
\end{equation}%
which has the solution 
\begin{equation}
N_{0}(x,t)=\frac{\kappa \sqrt{\gamma }M}{4}\left| x\right| +B(t)x+C(t)
\label{N0sol1pcp0}
\end{equation}%
However this solution cannot satisfy both the periodicity and smoothness
criteria for $M\neq 0$. \ Hence no solutions exist for $c_{+}^{2}=0$, and
there is no 1-particle version of the trumpet.

\bigskip \bigskip

\subsubsection{$c_{-}^{2}=-c^{2}>0$}

\bigskip

This case can easily be shown to be the analytic continution of the solution
for $c^{2}=c_{+}^{2}>0$, obtained by making the replacement $%
c_{+}^{2}\rightarrow ic_{-}^{2}$ in the solution (\ref{1psola}--\ref{1psolPi}%
). It is straightforward to check that, with this replacement, eqs. (\ref%
{N1p1}--\ref{N1p6}) are satisfied. \ \ 

However an important distinction arises in imposing periodicity on $%
N_{1}(x,t)$. This requirement now yields 
\begin{equation}
\cos (c_{-}L)+\frac{\kappa M\sqrt{\gamma }}{4c_{-}}\sin (c_{-}L)=\cos (L%
\sqrt{\gamma }\sqrt{\frac{\Lambda }{2}-\tau ^{2}})+\frac{\kappa M}{4\sqrt{%
\frac{\Lambda }{2}-\tau ^{2}}}\sin (L\sqrt{\gamma }\sqrt{\frac{\Lambda }{2}%
-\tau ^{2}})=\xi  \label{cons1pneg}
\end{equation}%
which has a different parameter space of solutions than eq. (\ref{cons1p}).
\ The solution is given by finding the intersection of the curve $\cos (\chi
)+m(\tau )\sin (\chi )$, (where now $\chi (\tau )=L\sqrt{\gamma }\sqrt{\frac{%
\Lambda }{2}-\tau ^{2}}>0$ and $m(\tau )=\frac{\kappa M}{4\sqrt{\frac{%
\Lambda }{2}-\tau ^{2}}}$) with a horizontal line of height $\xi $. The
left-hand side of (\ref{cons1pneg}) is never larger than $\sqrt{1+m^{2}(\tau
)}$, which attains its minimum at $\tau =0$. Hence only for $\left| \xi
\right| <\sqrt{1+\frac{\left( \kappa M\right) ^{2}}{8\Lambda }}$do
admissible solutions exist for the allowed range of \ $\tau $.\ \ 

Provided $\xi $ respects this bound, there will be a countably infinite set
of solutions for $\sqrt{\gamma }$, each parametrized by $\xi $. \ These are 
\begin{equation}
\sqrt{\gamma }=\frac{4m(\tau )}{\kappa ML}\left( \arctan \left( \frac{\xi 
\sqrt{m^{2}(\tau )+(1-\xi ^{2})}-m(\tau )}{m^{2}(\tau )-\xi ^{2}}\right)
+n\pi \right)  \label{gam1pmtautrig}
\end{equation}%
where $n$ is a positive integer for either choice of sign, and could also be
zero if the positive root is chosen. More explicitly, 
\begin{equation}
\sqrt{\gamma }=\frac{1}{L\sqrt{\Lambda /2-\tau ^{2}}}\left[ \arctan \left( 
\frac{\xi \sqrt{\frac{\kappa ^{2}M^{2}}{16}+(1-\xi ^{2})(\Lambda /2-\tau
^{2})}-\frac{\kappa M}{4}}{\frac{\kappa ^{2}M^{2}}{16}-\xi ^{2}(\Lambda
/2-\tau ^{2})}\sqrt{\Lambda /2-\tau ^{2}}\right) +n\pi \right]
\label{gam1psoltrig}
\end{equation}%
\ and 
\begin{equation}
H=\frac{4\dot{\tau}}{\kappa \sqrt{\tau ^{2}-\frac{\Lambda }{2}}}\left[
\arctan \left( \frac{\xi \sqrt{\frac{\kappa ^{2}M^{2}}{16}+(1-\xi
^{2})(\Lambda /2-\tau ^{2})}-\frac{\kappa M}{4}}{\frac{\kappa ^{2}M^{2}}{16}%
-\xi ^{2}(\Lambda /2-\tau ^{2})}\sqrt{\Lambda /2-\tau ^{2}}\right) +n\pi %
\right]  \label{Ham1ptrig}
\end{equation}%
which is the general form for the Hamiltonian for this case.

\bigskip

Although $\xi $ can take on both positive and negative values, the
requirement that $\sqrt{\gamma }>0$ must be maintained. \ For $n>0$ there
are no additional constraints since the $\arctan $ function is never less
than $-\pi /2$. \ However for $n=0,$ the argument of the $\arctan $ must be
positive, implying that $\xi >1$. This is because the left-hand side of (\ref%
{cons1pneg}) has a positive slope and a value of unity for $c_{-}L=0$, and
so the nearest positive root must have $\xi >1$. Hence solutions exist for $%
n=0$ provided 
\begin{equation}
\sqrt{1+\frac{\left( \kappa M\right) ^{2}}{8\Lambda }}>\xi >1
\label{xiconstraint}
\end{equation}%
holds. For $n>0$, only the left-hand inequality must be respected.

\bigskip

The single particle solutions for $c^{2}<0$ are therefore%
\begin{eqnarray}
&&  \nonumber \\
N_{0}(x,t) &=&-\frac{\dot{\tau}}{(\Lambda /2-\tau ^{2})}\left\{ 
\frac{{\mbox{sgn}}(\xi )\kappa M\cos (\left[ \tan ^{-1}\left( \sigma \left( \tau \right)
\right) +n\pi \right] (|x|/L-1))}{4\sqrt{\frac{\kappa ^{2}M^{2}}{16}+(1-\xi
^{2})(\tau ^{2}-\Lambda /2)}}-1\right\}  \label{1psolatrig} \\
N_{1}(x,t) &=&\frac{{\mbox{sgn}}(\xi )\kappa M{\mbox{sgn}}(x)\left[ \tan
^{-1}\left( \sigma \left( \tau \right) \right) +n\pi \right] \tau \dot{\tau}%
}{4L(\Lambda /2-\tau ^{2})^{2}\sqrt{\frac{\kappa ^{2}M^{2}}{16}+(1-\xi
^{2})(\Lambda /2-\tau ^{2})}}  \nonumber \\
&&\times \left[ \sin (\left[ \tan ^{-1}\left( \sigma \left( \tau \right)
\right) +n\pi \right] (|x|/L-1))-\frac{\sigma \left( \tau \right) \left(
-1\right) ^{n}(|x|/L-1)}{\sqrt{1+\sigma ^{2}\left( \tau \right) }}\right]
\label{1psolbtrig} \\
\Psi (x,t) &=&-2\ln \left\{ \frac{{\mbox{sgn}}(\xi )\kappa M\cos (\left[ \tan
^{-1}\left( \sigma \left( \tau \right) \right) +n\pi \right] (|x|/L-1))}{4%
\sqrt{\frac{\kappa ^{2}M^{2}}{16}+(1-\xi ^{2})(\Lambda /2-\tau ^{2})}}%
-1\right\}  \nonumber \\
&&-2\ln \left( \frac{{\frak L}^{2}\left[ \tau \pm \sqrt{\frac{\kappa
^{2}M^{2}}{16(1-\xi ^{2})}+(\Lambda /2-\tau ^{2})}\right] }{(\Lambda /2-\tau
^{2})}\sqrt{\frac{\kappa ^{2}M^{2}}{16(1-\xi ^{2})}+(\Lambda /2-\tau ^{2})}%
\right)  \label{1psolctrig} \\
\kappa \Pi &=&\frac{\pm \sqrt{\left| 1-\xi ^{2}\right| }\sqrt{(\Lambda
/2-\tau ^{2})}\left[ \tan ^{-1}\left( \sigma \left( \tau \right) \right)
+n\pi \right] }{\frac{\kappa M}{4}\cos (\left[ \tan ^{-1}\left( \sigma
\left( \tau \right) \right) +n\pi \right] (|x|/L-1))-\sqrt{\frac{\kappa
^{2}M^{2}}{16}+(1-\xi ^{2})(\Lambda /2-\tau ^{2})}}  \label{1psolPitrig}
\end{eqnarray}%
where now $\sigma \left( \tau \right) =\frac{s\xi \sqrt{\frac{\left( \kappa
M\ell \right) ^{2}}{16}+(1-\xi ^{2})(\Lambda /2-\tau ^{2})}-\frac{\kappa
M\ell }{4}}{\frac{\left( \kappa M\ell \right) ^{2}}{16}-\xi ^{2}(\Lambda
/2-\tau ^{2})}\sqrt{\Lambda /2-\tau ^{2}}$ . \ 

\bigskip Finally, there is a solution with $\dot{\tau}=0$. It is
straightforwardly obtained from eqs. (\ref{N1p1}--\ref{N1p6}):%
\begin{eqnarray}
N_{0}(x,t) &=&\hat{N}\cos (\left[ \tan ^{-1}\left( \frac{\kappa M\ell }{4}%
\right) +n\pi \right] (|x|/L-1))  \label{1pstat1} \\
N_{1}(x,t) &=&0  \label{1pstat2} \\
\Psi (x,t) &=&-2\ln \left\{ \hat{N}\cos (\left[ \tan ^{-1}\left( \frac{%
\kappa M\ell }{4}\right) +n\pi \right] (|x|/L-1))\right\} -2\hat{N}\frac{t}{%
\ell }  \label{1pstat3} \\
\kappa \Pi &=&\frac{\pm \tan ^{-1}\left( \frac{\kappa M\ell }{4}\right)
+n\pi }{\cos (\left[ \tan ^{-1}\left( \frac{\kappa M\ell }{4}\right) +n\pi %
\right] (|x|/L-1))}  \label{1pstat4}
\end{eqnarray}%
where the constant $\hat{N}$\ is arbitrary, $\tau =0$, 
\begin{equation}
\sqrt{\gamma }=\frac{\ell }{L}\left[ \tan ^{-1}\left( \frac{\kappa M\ell }{4}%
\right) +n\pi \right]  \label{statmet}
\end{equation}%
and the Hamiltonian vanishes.

\subsection{Analysis of Single-particle solutions}

The solution given in (\ref{1pstat1}--\ref{1pstat4}), with metric (\ref%
{statmet}) is a static spacetime in which the gravitational attraction of
the point mass exactly balances the tendency toward cosmological expansion
induced by a positive $\Lambda $. The extrinsic curvature (\ref{tauK})
vanishes, and so cannot be used as a time coordinate. \ 

More generally the presence of the point mass will alter the
expansion/contraction of the spacetime, depending on the relative values of
the parameters. \ I shall consider four different choices for $\tau $,
classified according to the behaviours of the spacetime in the $M=0$ case.

\begin{enumerate}
\item $\tau =t/\ell ^{2}$. \ Here the extrinsic curvature is taken to be the
time coordinate itself; from (\ref{ham1}) the Hamiltonian is proportional to
the circumference of the circle.

\item Proper-time coordinates. In these coordinates $\tau $ is chosen so
that in the $M=0$ limit of the spacetime, $t$ is the proper time. Although
this no longer holds when $M\neq 0$, this (slightly abusive) terminology
will be retained.

\item $\tau =-1/t$. For this choice the circumference vanishes at $t=0$,
leading to a ``big bang'' expansion of the spacetime.

\item $\dot{\tau}\ell =\sqrt{\pm \left( \tau ^{2}-\Lambda /2\right) }$. \
For this choice the Hamiltonian is constant throughout the evolution when $%
M=0$.
\end{enumerate}

\bigskip

Although these choices are all locally equivalent, globally they cannot be
transformed into each other at points where the spatial metric either
diverges or vanishes. \ 

\bigskip

\subsubsection{$\Lambda <0$}

For $\Lambda <0$ the circle expands from zero to some maximal size and then
recontracts. \ The attractive gravitational effects of the point mass reduce
the maximal size attained in the expansion. \ Setting $\tau =t/\ell ^{2}$
and $\left| \Lambda \right| =2/\ell ^{2}$, the proper circumference ${\frak C%
}$, of the circle is 
\begin{equation}
{\frak C}=\frac{2\ell }{\sqrt{\left( t/\ell \right) ^{2}+1}}\text{arctanh}%
\left( \frac{\xi \sqrt{\frac{\left( \kappa M\ell \right) ^{2}}{16}+(\xi
^{2}-1)(\left( t/\ell \right) ^{2}+1)}-\frac{\kappa M\ell }{4}}{\frac{\left(
\kappa M\ell \right) ^{2}}{16}+\xi ^{2}(\left( t/\ell \right) ^{2}+1)}\sqrt{%
\left( t/\ell \right) ^{2}+1}\right)  \label{an1}
\end{equation}%
and the accompanying Hamiltonian $H=\frac{2{\frak C}}{\kappa \ell ^{2}}$ for
this choice of time coordinate. The circumference has a value of $2\ell $
arccosh$\xi $ \ at $\tau =0$ when $M=0$.

\bigskip

\begin{figure}[tbp]
\begin{center}
\epsfig{file=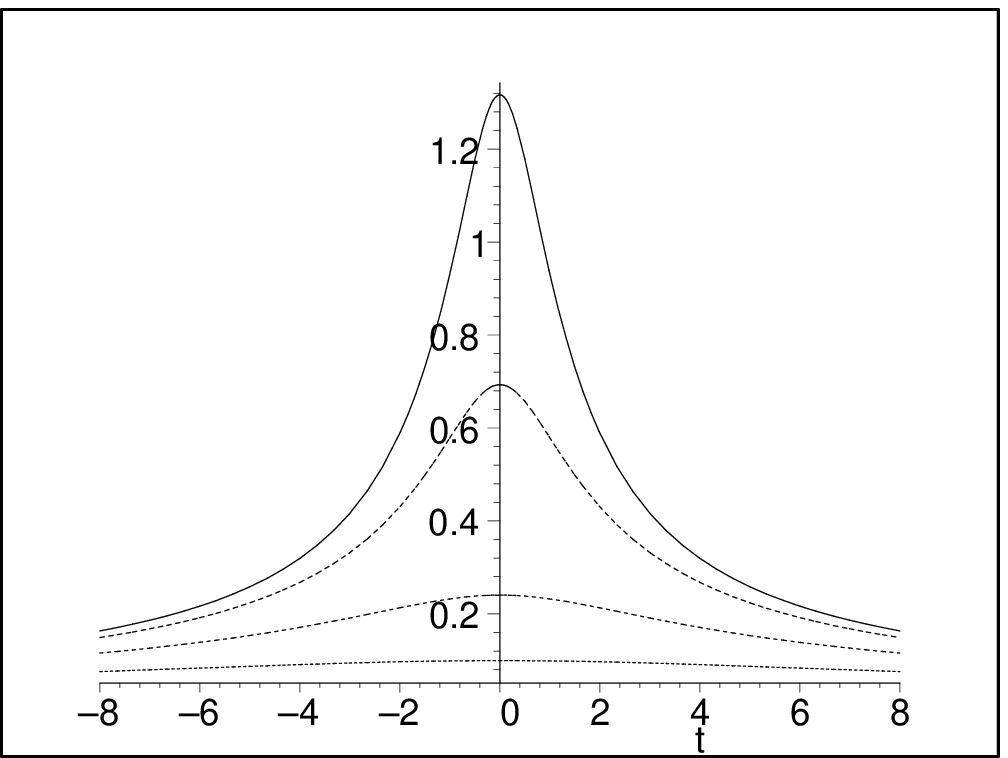,width=0.8\linewidth}
\end{center}
\caption{Circumference plotted against time for $\protect\xi =2$ for
differing values of ${\cal M}=0$ (solid), $1$ (hash), $4 $ (dash), $10$
(dot), with $\Lambda <0$.}
\label{fig1}
\end{figure}

\begin{figure}[tbp]
\begin{center}
\epsfig{file=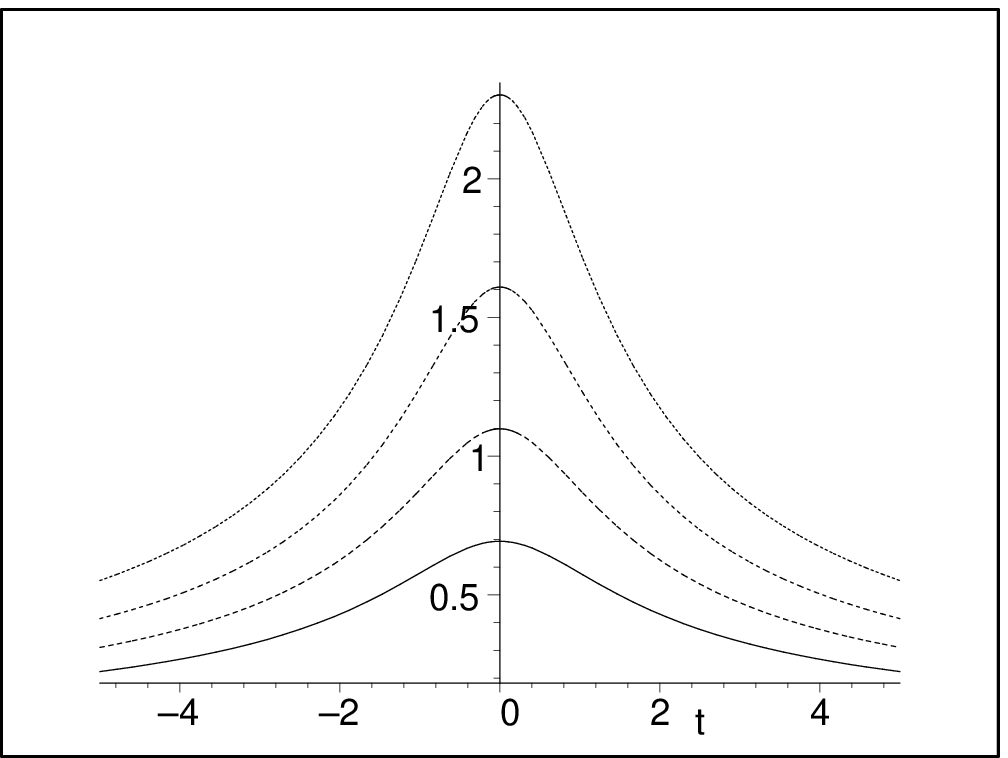,width=0.8\linewidth}
\end{center}
\caption{Circumference plotted against time for ${\cal M}=1$ for differing
values of $\protect\xi =2$ (solid), $3$ (hash), $5$ (dash), $10$ (dot), with 
$\Lambda <0$.}
\label{fig2}
\end{figure}

\begin{figure}[tbp]
\begin{center}
\epsfig{file=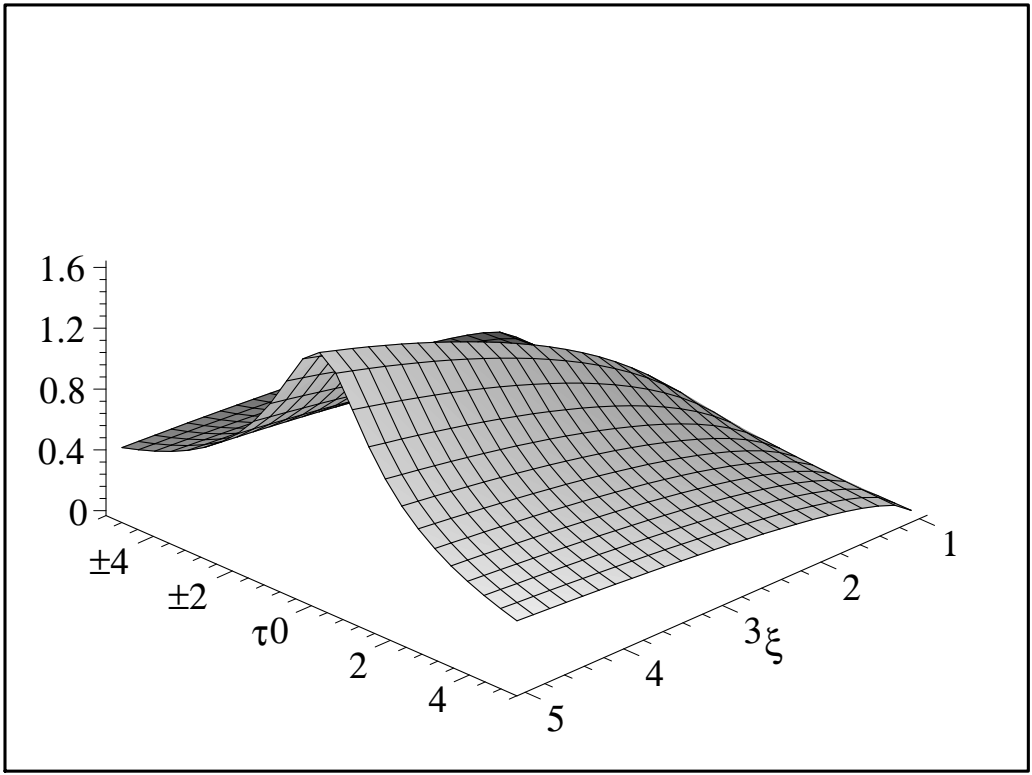,width=0.8\linewidth}
\end{center}
\caption{The $(t/\ell ,\protect\xi )$ circumference surface, for ${\cal M}=1$%
, , with $\Lambda <0$.}
\label{fig3}
\end{figure}

\begin{figure}[tbp]
\begin{center}
\epsfig{file=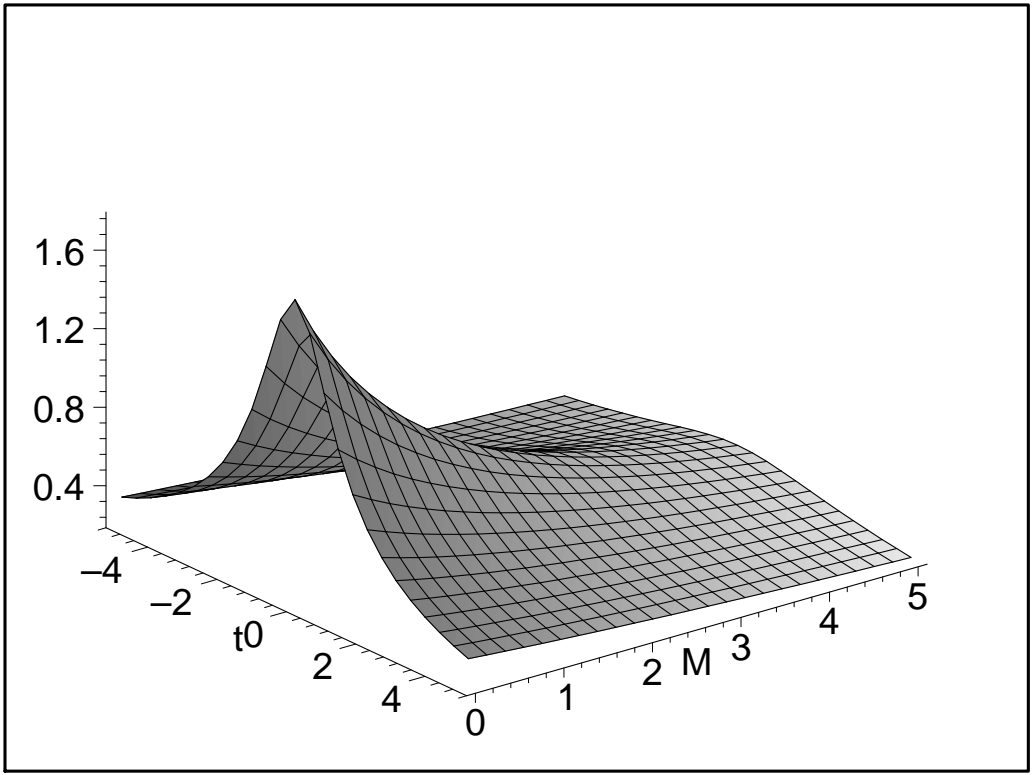,width=0.8\linewidth}
\end{center}
\caption{The $(t/\ell ,{\cal M})$ circumference surface, for $\protect\xi =3$%
, , with $\Lambda <0$.}
\label{fig4}
\end{figure}

Setting $\frac{\kappa M\ell }{4}={\cal M}$ ,\ figures \ref{fig1} -- \ref%
{fig4} plot the circumference (in units of $2\ell $) against $t/\ell $ for
various values of ${\cal M}$ and $\xi $. The maximal expansion of the circle
decreases for increasing ${\cal M}$ (figs. \ref{fig1} and \ref{fig3}) and
increases for increasing $\xi $ (figs. \ref{fig2} and \ref{fig4}). \ 

\bigskip 

\bigskip 

\bigskip 

\bigskip 

\begin{figure}[tbp]
\begin{center}
\epsfig{file=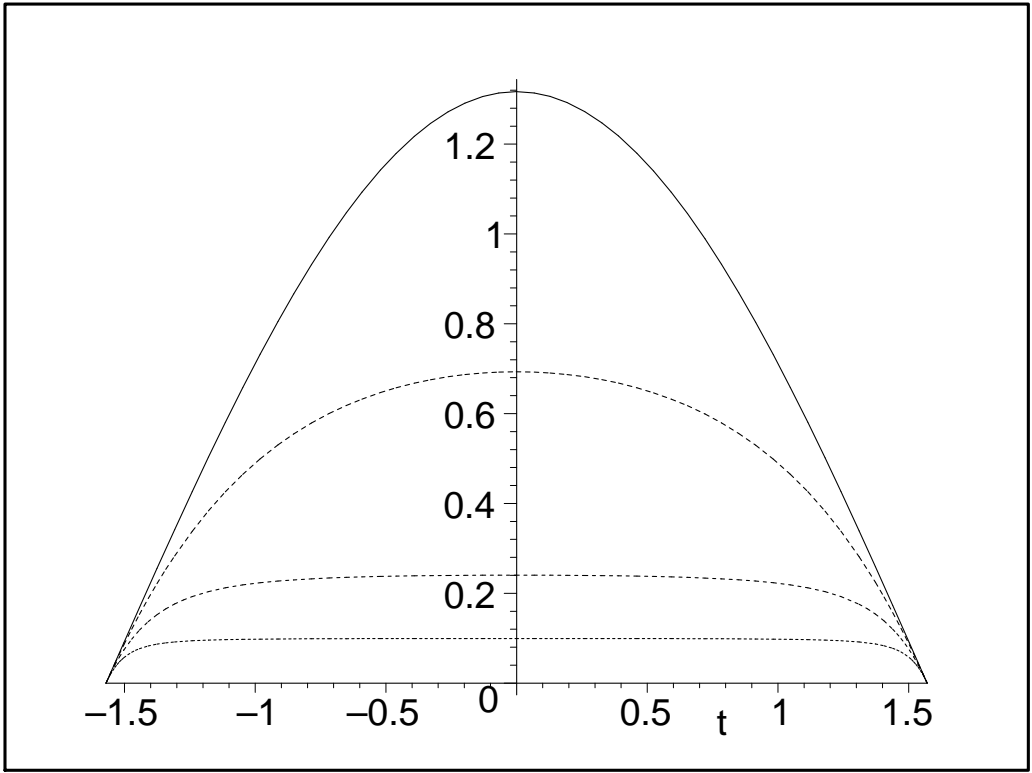,width=0.8\linewidth}
\end{center}
\caption{The same plot as in fig \ref{fig1} but in proper time coordinates $%
\protect\tau =\frac{1}{\ell }\tan \left( \frac{t}{\ell }\right) $, with $%
\Lambda <0$.}
\label{fig5}
\end{figure}

\begin{figure}[tbp]
\begin{center}
\epsfig{file=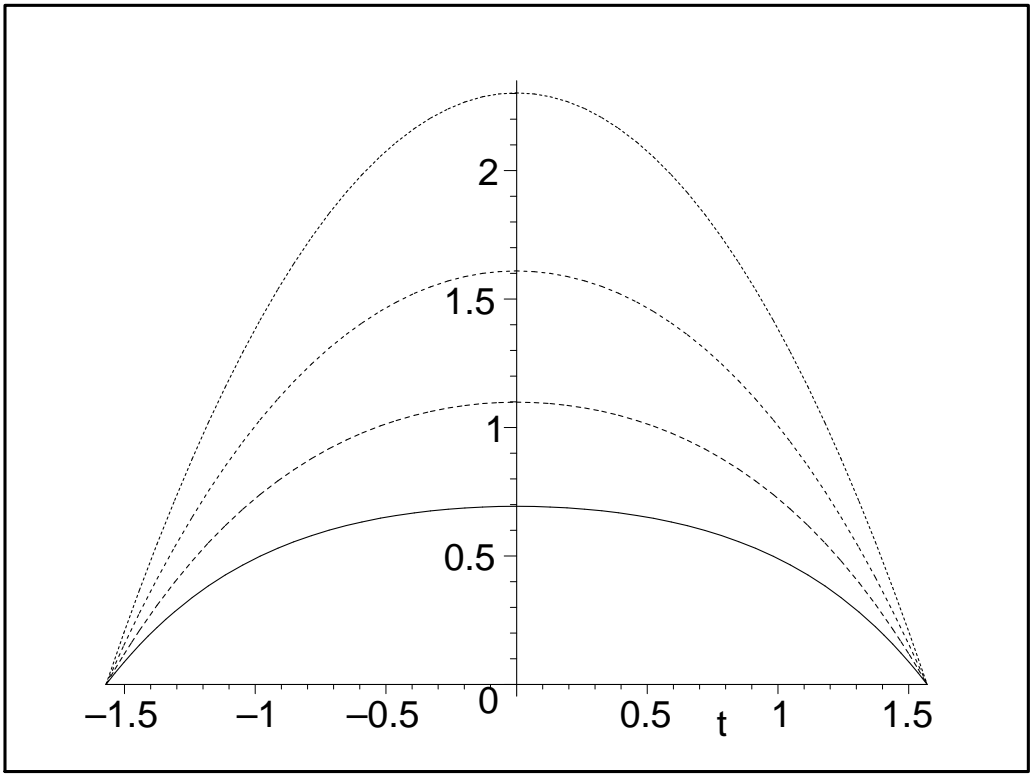,width=0.8\linewidth}
\end{center}
\caption{The same plot as in fig \ref{fig2} but with $\protect\tau =\frac{1}{%
\ell }\tan \left( \frac{t}{\ell }\right) $, with $\Lambda <0$.}
\label{fig6}
\end{figure}

Another useful way of understanding the effect of the point mass on the
spacetime is to choose $\tau =\frac{1}{\ell }\tan \left( \frac{t}{\ell }%
\right) $ as in eq. (\ref{adstau}). In these coordinates $t$ is the proper
time when $M=0$. This gives%
\begin{equation}
{\frak C}=2\ell \cos \left( \frac{t}{\ell }\right) \text{arctanh}\left( 
\frac{\xi \sqrt{\frac{\left( \kappa M\ell \right) ^{2}}{16}\cos ^{2}\left( 
\frac{t}{\ell }\right) +(\xi ^{2}-1)}-\frac{\kappa M\ell }{4}\cos \left( 
\frac{t}{\ell }\right) }{\frac{\left( \kappa M\ell \right) ^{2}}{16}\cos
^{2}\left( \frac{t}{\ell }\right) +\xi ^{2}}\right)  \label{an2}
\end{equation}%
where now $H=\frac{2{\frak C}}{\kappa \ell ^{2}}\sec ^{2}\left( t/\ell
\right) $, and $t\in \left[ -\frac{\pi \ell }{2},\frac{\pi \ell }{2}\right] $%
. \ Figures \ref{fig5} and \ref{fig6} show the evolution of the
circumference over the full range of $t$ for various values of ${\cal M}$
and $\xi $ respectively. As the mass increases, the circumference of the
circle more rapidly approaches its decreasing maximal value, hovering there
for most of the evolution of the spacetime. \bigskip 
\bigskip 

By replacing $t\rightarrow \ell ^{2}/t$ in (\ref{an1}) (so that $\tau =-1/t$%
) a ``big bang'' cosmology is obtained, in which the Hamiltonian is
divergent at $t=0$ and the spacetime expands from that point. In these
coordinates, the evolution of the circumference is illustrated in figure \ref%
{fig1a} \bigskip 
\begin{figure}[tbp]
\begin{center}
\epsfig{file=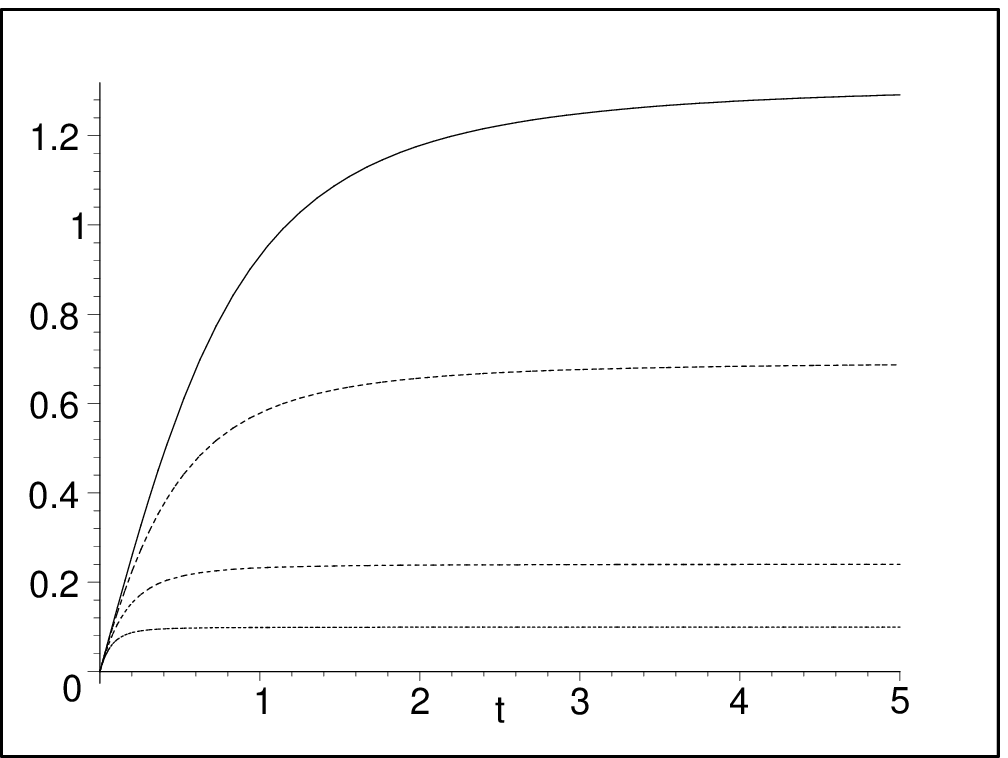,width=0.8\linewidth}
\end{center}
\caption{Circumference plotted against time in the ``big bang'' coordinates,
with $\Lambda <0$. Here $\protect\xi =2$ and the values of ${\cal M}$ are
respectively $0$ (solid), $1 $ (hash), $4$ (dash), $10$ (dot).}
\label{fig1a}
\end{figure}
with both the negative cosmological constant and the mass contributing to
the deceleration. For $M=0$ the circumference asymptotes to a value of unity
in units of $2\ell $ due to the decelerating effects of the negative
cosmological constant.

\begin{figure}[tbp]
\begin{center}
\epsfig{file=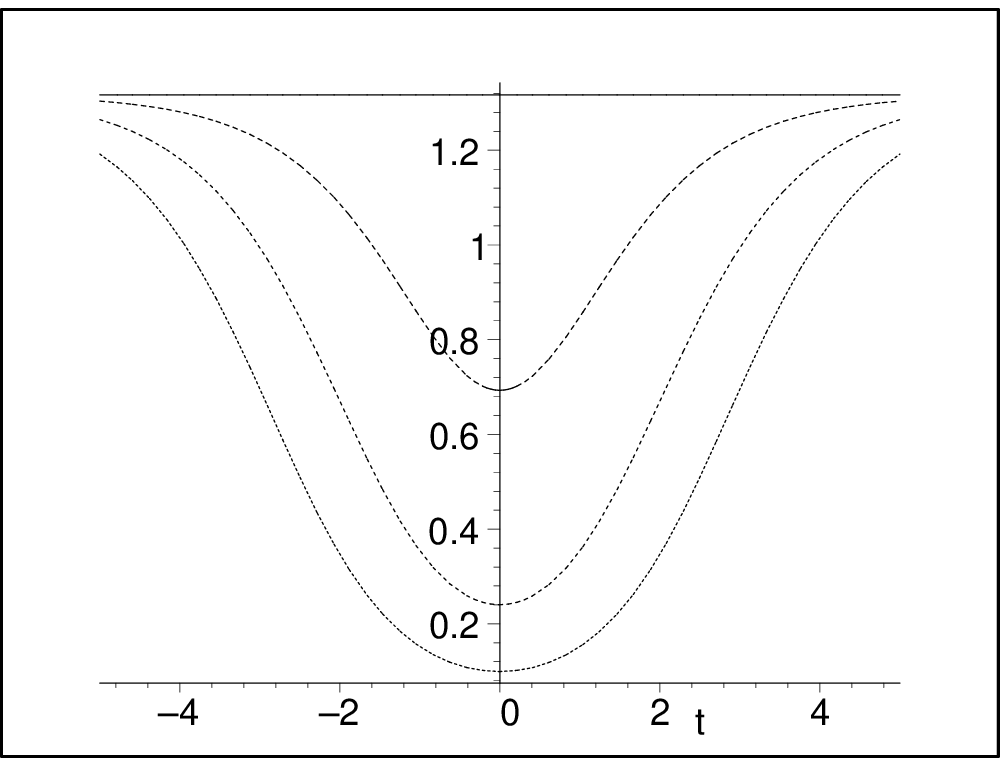,width=0.8\linewidth}
\end{center}
\caption{Hamiltonian plotted against time in conformally flat coordinates for%
$\ \protect\xi =2$ for differing values of ${\cal M}=0$ (solid), $1$ (hash), 
$4$ (dash), $10$ (dot), with $\Lambda <0$.}
\label{fig7}
\end{figure}

\begin{figure}[tbp]
\begin{center}
\epsfig{file=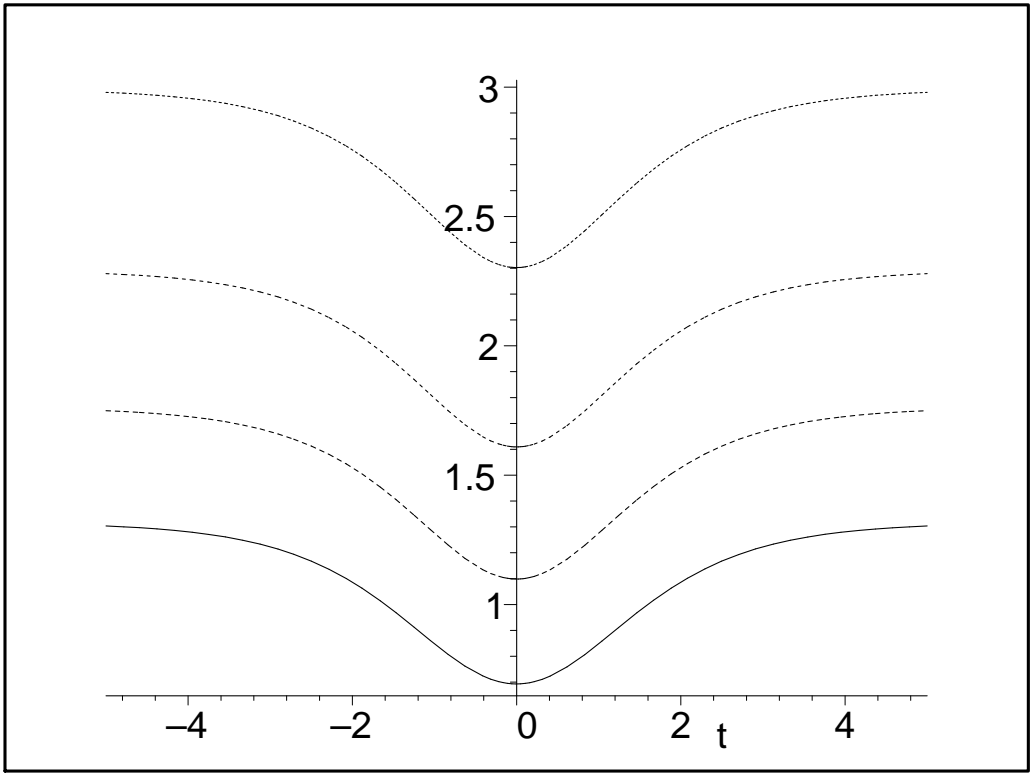,width=0.8\linewidth}
\end{center}
\caption{Hamiltonian plotted against time in conformally flat coordinates
for ${\cal M}=1$ for differing values of $\protect\xi =2$ (solid), $3$
(hash), $5$ (dash), $10$ (dot), with $\Lambda <0$.}
\label{fig8}
\end{figure}

The last choice of time coordinate I shall consider is that of the
conformally flat coordinates eq. (\ref{cplus3}), for which $\tau =\sinh
\left( t/\ell \right) /\ell $. This gives 
\begin{equation}
{\frak C}=2\ell \text{sech}\left( \frac{t}{\ell }\right) \text{arctanh}%
\left( \frac{\xi \sqrt{\frac{\left( \kappa M\ell \right) ^{2}}{16}\text{sech}%
^{2}\left( \frac{t}{\ell }\right) +(\xi ^{2}-1)}-\frac{\kappa M\ell }{4}%
\text{sech}\left( \frac{t}{\ell }\right) }{\frac{\left( \kappa M\ell \right)
^{2}}{16}\text{sech}^{2}\left( \frac{t}{\ell }\right) +\xi ^{2}}\right)
\label{an3}
\end{equation}%
for the circumference and $H=\frac{2{\frak C}}{\kappa \ell ^{2}}\cosh \left(
t/\ell \right) $ for the Hamiltonian. The Hamiltonian is constant for $M=0$%
,. and so in these coordinates departures from constant energy signal the
presence of \ a point mass. \ Figures \ \ref{fig7}\ and \ref{fig8}\ plot the
Hamiltonian\ in units of $\kappa \ell ^{2}/2$ for various values of \ ${\cal %
M}$ and $\xi $ respectively. \bigskip 
\bigskip 

\subsubsection{$\Lambda =0$}

For $\Lambda =0$ the circumference undergoes a perturally decelerating
expansion due to the presence of the point mass. Setting $\tau =t/\ell ^{2}$%
, the Hamiltonian is again $H=\frac{2{\frak C}}{\kappa \ell ^{2}}$ and 
\begin{equation}
{\frak C}=\frac{2\ell ^{2}}{\left| t\right| }\text{arctanh}\left( \frac{\xi 
\sqrt{\frac{\left( \kappa M\ell \right) ^{2}}{16}+(\xi ^{2}-1)\left( t/\ell
\right) ^{2}}-\frac{\kappa M\ell }{4}}{\frac{\left( \kappa M\ell \right) ^{2}%
}{16}+\xi ^{2}\left( t/\ell \right) ^{2}}\left| t/\ell \right| \right)
\label{an4}
\end{equation}%
where the spatial coordinate $x$ has been rescaled as before and $\ell $ is
now an arbitrary constant. In these coordinates the evolution of the
spacetime is qualitatively similar to that given in figures \ref{fig1} -- %
\ref{fig4}: the circumference expands from zero radius at $t=-\infty $ to
some maximal value and then reverses its evolution. This maximal value
increases as the mass decreases, diverging at $M\rightarrow 0$, in which
case the curves bifurcates into two distinct spactime evolutions, one
expanding to infinity and one contracting from infinity, that are related to
one another under $t\leftrightarrow -t$. \ 

\begin{figure}[tbp]
\begin{center}
\epsfig{file=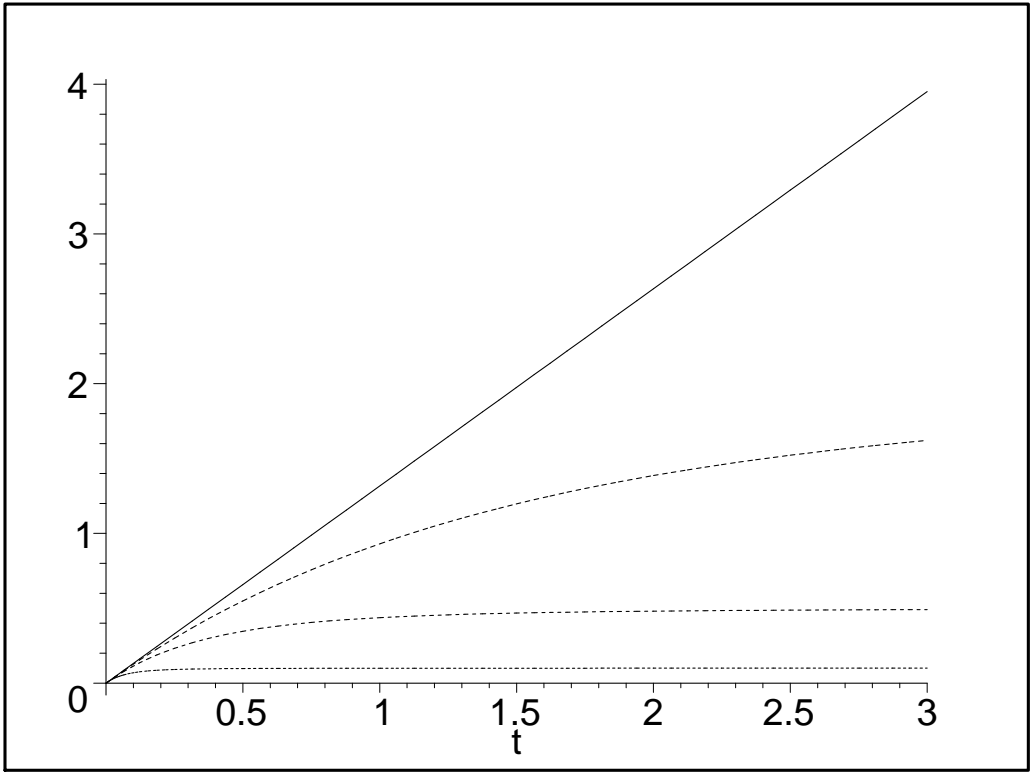,width=0.8\linewidth}
\end{center}
\caption{Circumference plotted against time for $\protect\xi =2$ for
differing values of ${\cal M}=0$ (solid), $1/2$ (hash), $2$ (dash), $10$
(dot), with $\Lambda =0$.}
\label{fig9}
\end{figure}

\begin{figure}[tbp]
\begin{center}
\epsfig{file=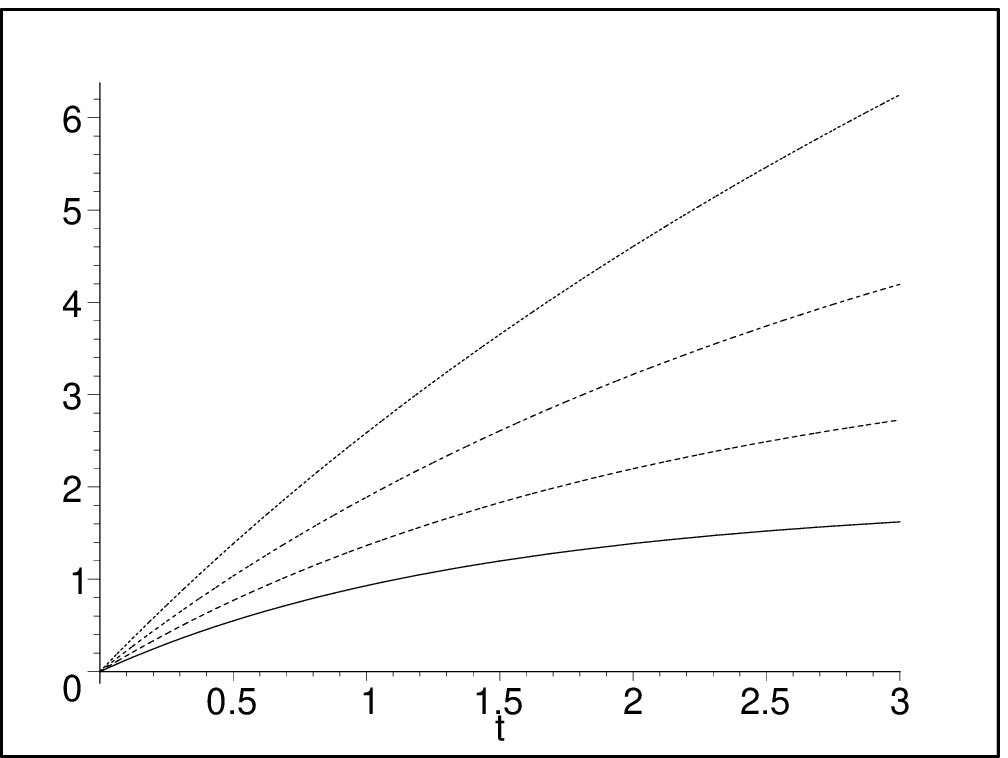,width=0.8\linewidth}
\end{center}
\caption{Circumference plotted against time for ${\cal M}=1/2$ for differing
values of $\protect\xi =2$ (solid), $3$ (hash), $5$ (dash), $10$ (dot), with 
$\Lambda =0$.}
\label{fig10}
\end{figure}

A more useful comparison is made in coordinates where $\tau =-1/t$, for
which $t$ is now the proper time when $M=0$. The circumference becomes 
\begin{equation}
{\frak C}=2\left| t\right| \text{arctanh}\left( \frac{\xi \sqrt{\frac{\left(
\kappa M\ell \right) ^{2}}{16}\left( t/\ell \right) ^{2}+(\xi ^{2}-1)}-\frac{%
\kappa M\ell }{4}\left| t/\ell \right| }{\frac{\left( \kappa M\ell \right)
^{2}}{16}\left( t/\ell \right) ^{2}+\xi ^{2}}\right)  \label{an4a}
\end{equation}%
and the Hamiltonian $H=\frac{2{\frak C}}{\kappa t^{2}}$ , which diverges at $%
t=0$. \ Figures \ref{fig9} and \ref{fig10} illustrate how the expansion of
the circumference is altered by the point mass in these coordinates. For $%
M=0 $ it expands indefinitely from a ``big bang'' as a linear function of $t$%
, whereas for $M\neq 0$ \ it asymptotes to 
\begin{equation}
\lim_{t\rightarrow \infty }{\frak C=}\frac{8(\xi -1)}{\kappa M\text{ }}
\label{an5}
\end{equation}%
approaching this value as $1/t^{2}$. The rate of deceleration is decreased
for increasing $\xi $. \bigskip 
\bigskip 
This is similar to the expansion in the big-bang coordinates for the $%
\Lambda <0$ case, except that for $M=0$ there is no deceleration.

\begin{figure}[tbp]
\begin{center}
\epsfig{file=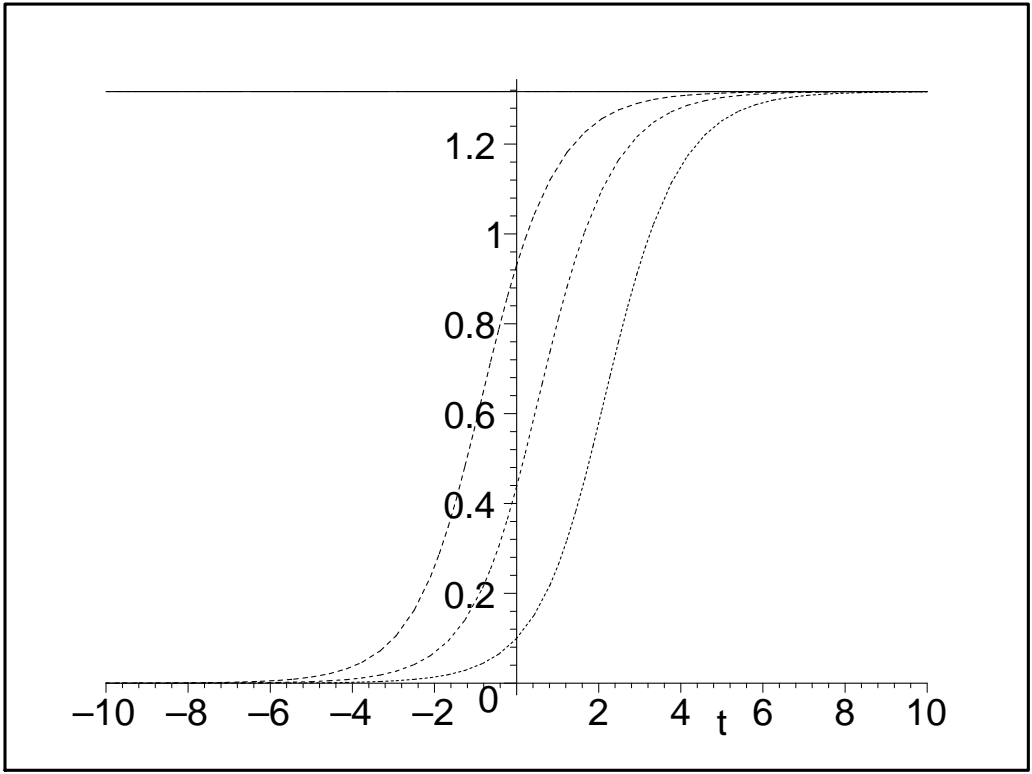,width=0.8\linewidth}
\end{center}
\caption{Hamiltonian plotted against time in conformally flat coordinates for%
$\ \protect\xi =2$ for differing values of ${\cal M}=0$ (solid), $1$ (hash), 
$4$ (dash), $10$ (dot), with $\Lambda =0$.}
\label{fig11}
\end{figure}

In coordinates which are conformally flat for $M=0$, the circumference is 
\begin{equation}
{\frak C}=2\ell \exp \left( -t/\ell \right) \text{arctanh}\left( \frac{\xi 
\sqrt{\frac{\left( \kappa M\ell \right) ^{2}}{16}+(\xi ^{2}-1)e^{2t/\ell }}-%
\frac{\kappa M\ell }{4}}{\frac{\left( \kappa M\ell \right) ^{2}}{16}+\xi
^{2}e^{2t/\ell }}e^{t/\ell }\right)  \label{an6}
\end{equation}%
with $H=\frac{2{\frak C}}{\kappa t^{2}}\exp \left( t/\ell \right) $, which
is constant. Figure \ref{fig11}\ \ plots the time dependence of the
Hamiltonian in these coordinates for various values of ${\cal M}$. \bigskip 
Except for $M=0$, the Hamiltonian vanishes as $t\rightarrow -\infty $, and
changes to a value of $\frac{2\text{arccosh}\xi }{\kappa t}$ around $t=0$,
exponentially rapidly approaching this value for large $t.$

\subsubsection{$\Lambda >0$}

For $\Lambda >0$ a wide variety of possibilities emerge for the evolution of
the circumference. This is because the attractive gravitational character of
the mass is offset by the `repulsive' gravitational character of the
positive cosmological constant. \ 

\begin{figure}[tbp]
\begin{center}
\epsfig{file=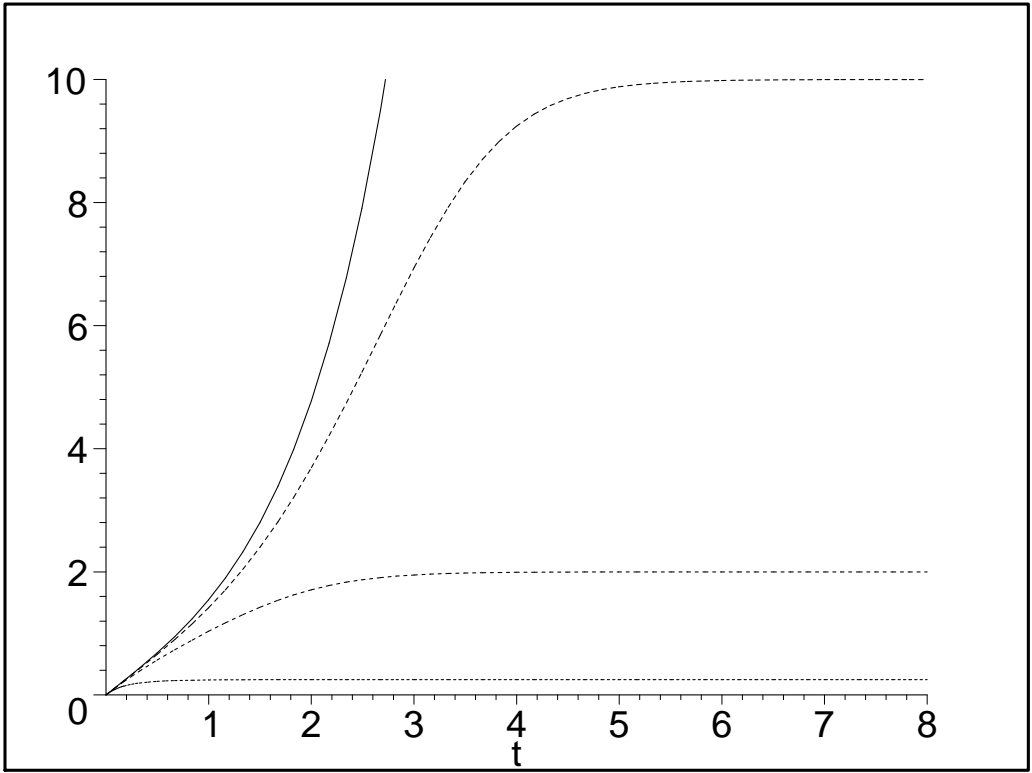,width=0.8\linewidth}
\end{center}
\caption{Circumference plotted against the proper time choice for $\protect%
\xi =2$ for differing values of ${\cal M}=0$ (solid), $0.1$ (hash), $1/2$
(dash), $4$ (dot), with $\Lambda >0$.}
\label{fig12}
\end{figure}

Consider first the behaviour in the proper-time coordinates (\ref{tausol2}),
\ where $\tau =-\coth \left( t/\ell \right) /\ell .$ From (\ref{gam1psol})
this gives for the circumference%
\begin{equation}
{\frak C}=2\ell \sinh \left( \frac{t}{\ell }\right) \text{arctanh}\left( 
\frac{\xi \sqrt{\frac{\left( \kappa M\ell \right) ^{2}}{16}\sinh ^{2}\left( 
\frac{t}{\ell }\right) +(\xi ^{2}-1)}-\frac{\kappa M\ell }{4}\sinh \left( 
\frac{t}{\ell }\right) }{\frac{\left( \kappa M\ell \right) ^{2}}{16}\sinh
^{2}\left( \frac{t}{\ell }\right) +\xi ^{2}}\right)  \label{an9}
\end{equation}%
where $H=\frac{2{\frak C}}{\kappa \ell ^{2}}$csch$^{2}\left( t/\ell \right) $%
. For $M=0$ the spacetime exponentially expands as in the dish scenario as
described by (\ref{dsmet2}). \ However for $M\neq 0$ the exponential
expansion halts for sufficiently large $t$, and the circumference asymptotes
exponentially to the fixed value \ (\ref{an5}), as shown in fig. \ref{fig12}%
.\bigskip 

\begin{figure}[tbp]
\begin{center}
\epsfig{file=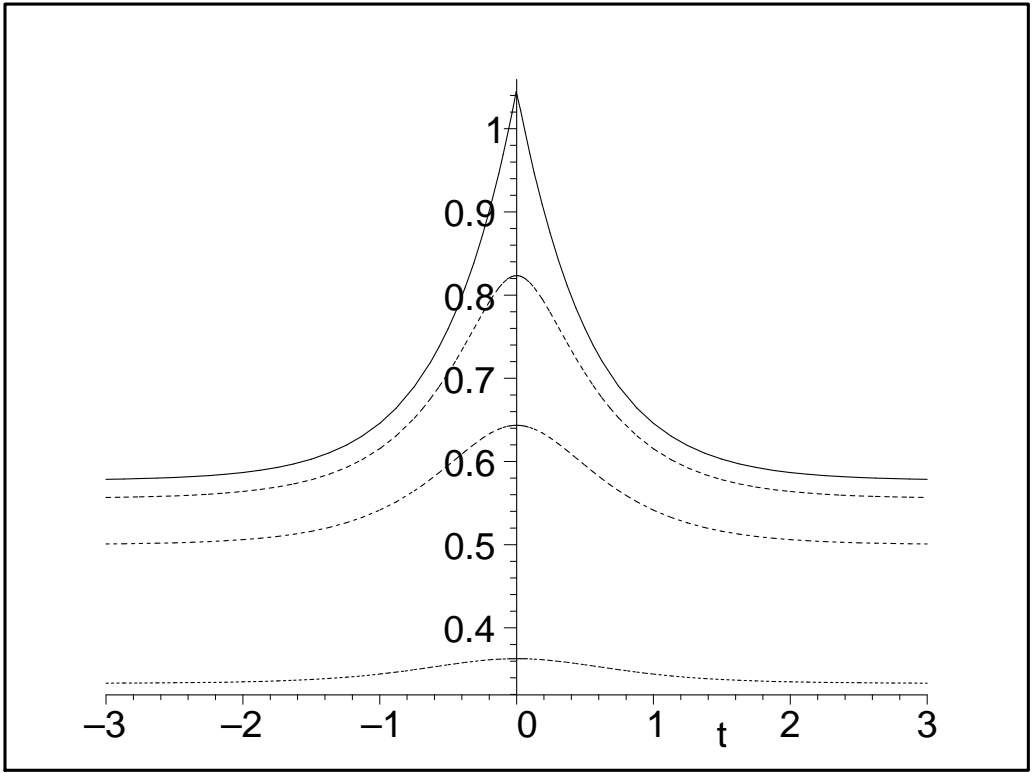,width=0.8\linewidth}
\end{center}
\caption{Circumference plotted against the proper time choice for $\protect%
\xi =2$ for differing values of ${\cal M}=\protect\sqrt{3}$ (solid), $1.8$
(hash), $2$ (dash), $3$ (dot), with $\Lambda >0$, from eq. (\ref{an10}) . \
Note the presence of the cusp for ${\cal M}=\protect\sqrt{3}$. }
\label{fig13}
\end{figure}

For the generalization of the candlestick scenario (\ref{dsmet1}), a wider
variety of possibilities ensues. \ The proper time coordinate choice is now $%
\tau =\tanh \left( t/\ell \right) /\ell $, yielding 
\begin{equation}
{\frak C}=2\ell \cosh \left( \frac{t}{\ell }\right) \left[ \arctan \left( 
\frac{\xi \sqrt{\frac{\left( \kappa M\ell \right) ^{2}}{16}\cosh ^{2}\left( 
\frac{t}{\ell }\right) +(1-\xi ^{2})}-\frac{\kappa M\ell }{4}\cosh \left( 
\frac{t}{\ell }\right) }{\frac{\left( \kappa M\ell \right) ^{2}}{16}\cosh
^{2}\left( \frac{t}{\ell }\right) -\xi ^{2}}\right) +n\pi \right]
\label{an10}
\end{equation}%
The Hamiltonian $H=\frac{2{\frak C}}{\kappa \ell ^{2}}$sech$^{2}\left(
t/\ell \right) $ and does not diverge anywhere. \ \ The spacetime \ behaves
completely differently for $n=0$ than for the candlestick metric (\ref%
{dsmet1}): the circumference evolves from the asymptotic value $\frac{8(\xi
-1)}{\kappa M\text{ }}$ at $t=-\infty $, grows to some maximal size, and
then recontracts, reversing its evolution to $t=+\infty $, \ When the
left-hand inequality of (\ref{xiconstraint}) is saturated, a cusp develops
at $t=0$, as illustrated in fig. \ref{fig13}. \bigskip 

\begin{figure}[tbp]
\begin{center}
\epsfig{file=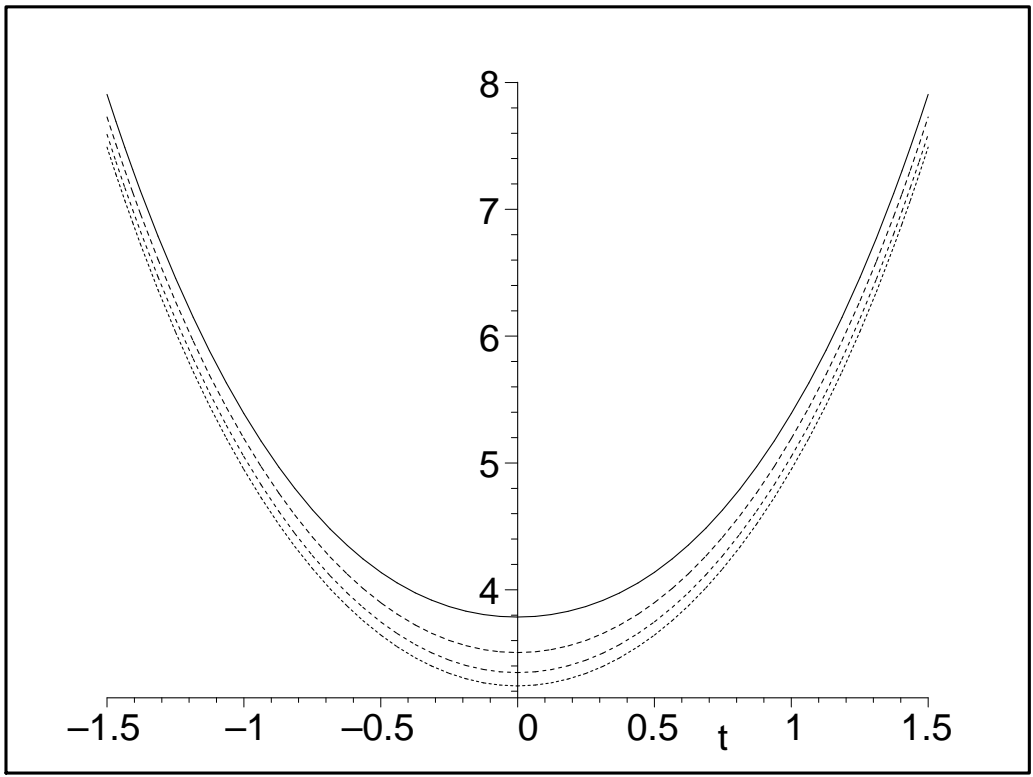,width=0.8\linewidth}
\end{center}
\caption{ Circumference plotted against the proper time choice for $\protect%
\xi =2$ for differing values of ${\cal M}=2$ (solid), $3$ (hash), $5$
(dash), $10$ (dot), with $\Lambda >0$, from eq. (\ref{an10}), with $n=1$ .}
\label{fig14}
\end{figure}

\begin{figure}[tbp]
\begin{center}
\epsfig{file=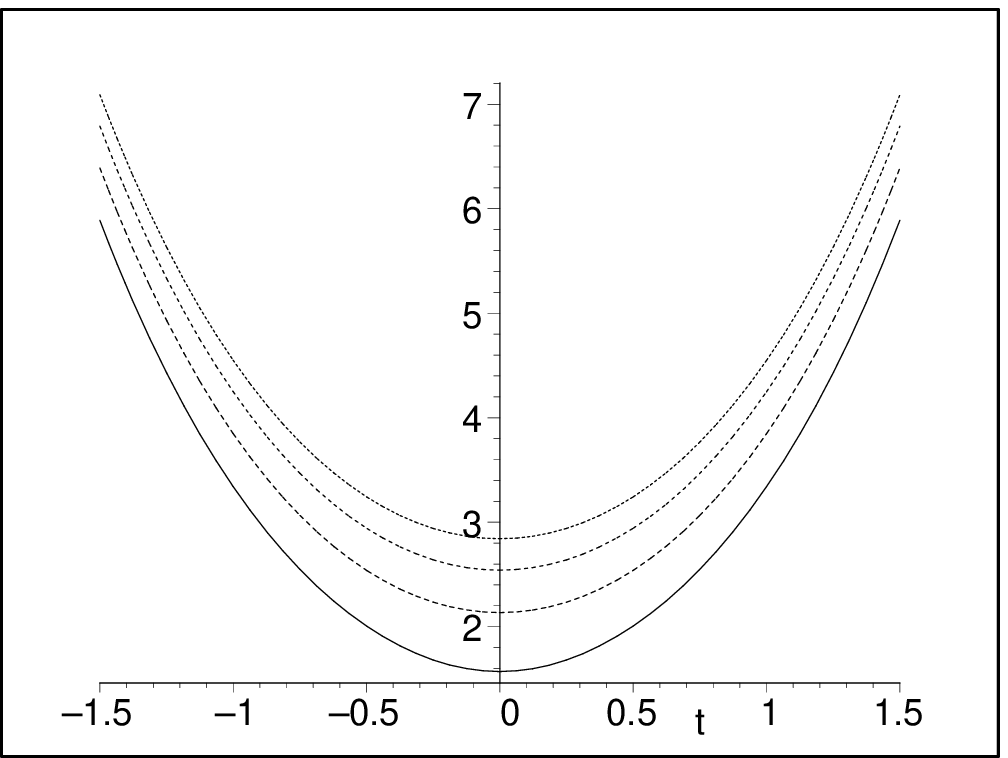,width=0.8\linewidth}
\end{center}
\caption{ The same situation as in fig \ref{fig14}, except that $\protect\xi %
=-2$. }
\label{fig15}
\end{figure}

For $n>0$, the only constraint on $\xi $ is $\sqrt{1+\frac{\left( \kappa
M\right) ^{2}}{8\Lambda }}>\left| \xi \right| $. \ The evolution is now
analogous to that of the metric (\ref{dsmet1}), with the circle having a
large initial circumference that exponentially shrinks with proper time to a
minimal value and then exponentially expands again to infinity, each
labelled by a positive integer $n$. \ Fig. \ref{fig14}\ shows a set of
typical values for $n=1$; larger values of $n$ follow a similar pattern, but
with the minimal value shifted up by $2\ell \left( n-1\right) \pi $. The
effect of the point mass is to reduce the size of the minimal circumference.
\bigskip 
If $\xi <0$, the effect of the point mass is to enlarge the size of the
minimal circumference; fig \ref{fig15} provides an illustration. \bigskip 

\begin{figure}[tbp]
\begin{center}
\epsfig{file=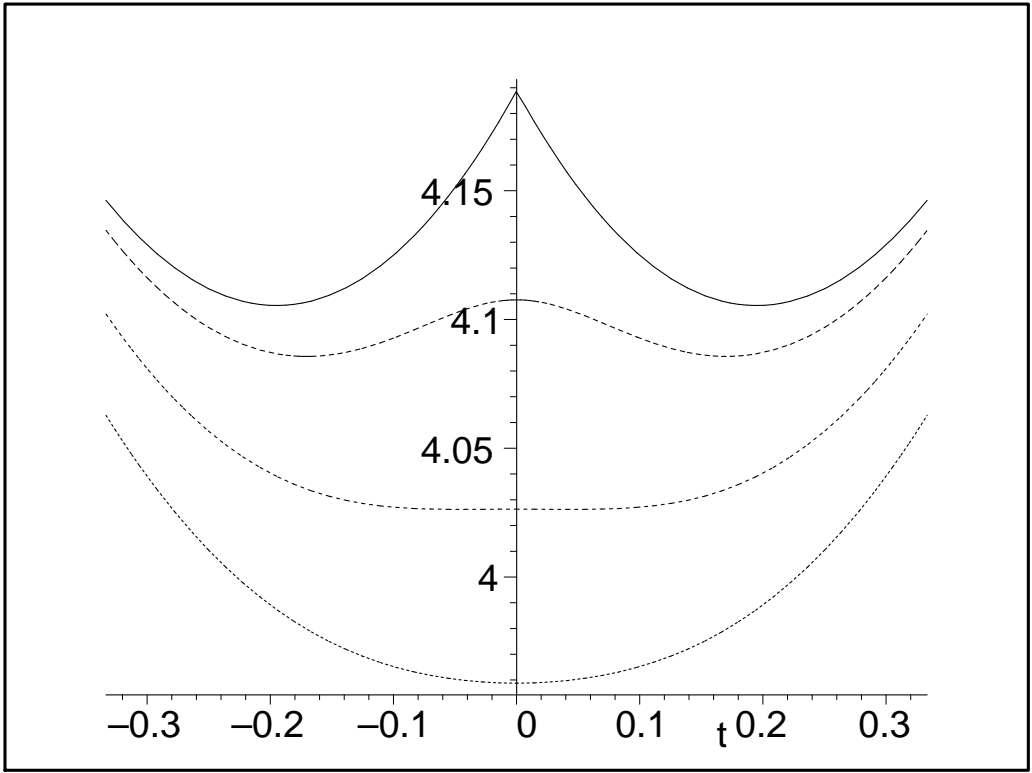,width=0.8\linewidth}
\end{center}
\caption{A close-up near $t=0$ of the evolution of the circumference in (\ref%
{an10}) for $\Lambda >0$ , $n=1$, and $\frac{\protect\kappa M\ell }{4}=%
\protect\sqrt{3}$ in proper time coordinates, for $\protect\xi =$ $\protect%
\sqrt{1+\frac{\left( \protect\kappa M\right) ^{2}}{8\Lambda }}$ (solid), $%
\protect\xi =\protect\sqrt{1+\frac{\left( \protect\kappa M\right) ^{2}}{%
8\Lambda }}\left( 1-\protect\varepsilon /4\right) $ (hash), $\protect\xi =%
\protect\sqrt{1+\frac{\left( \protect\kappa M\right) ^{2}}{8\Lambda }}\left(
1-\protect\varepsilon \right) $ (dash), $\protect\xi =\protect\sqrt{1+\frac{%
\left( \protect\kappa M\right) ^{2}}{8\Lambda }}\left( 1-2\protect%
\varepsilon \right) $ (dot), }
\label{fig16}
\end{figure}

For most values of $\xi $ there is a unique minimal circumference. However
for a certain range 
\begin{equation}
\sqrt{1+\frac{\left( \kappa M\ell \right) ^{2}}{16}}>\left| \xi \right| >%
\sqrt{1+\frac{\left( \kappa M\right) ^{2}}{8\Lambda }}\left( 1-\varepsilon
\right)  \label{an11}
\end{equation}%
where 
\begin{equation}
\varepsilon =\frac{\left( \frac{\kappa M\ell }{4}\right) ^{4}}{\left[ \frac{%
\kappa M\ell }{4}+\left( 1+\frac{\left( \kappa M\ell \right) ^{2}}{16}%
\right) \left( \arctan \left( \frac{\kappa M\ell }{4}\right) +n\pi \right) %
\right] }  \label{an12}
\end{equation}%
there is a local maximum in the evolution of the circumference at $t=0$. \
Beginning at $t=-\infty $, the circumference evolves from infinity to a
minimal value, expands to a local maximum at $t=0$, and then reverses its
evolution. \ If the left-hand bound given in (\ref{an11}) is saturated, a
cusp develops at $t=0$. \ Fig. \ref{fig16} illustrates the effect near $\xi
=2$ (i.e. for $\frac{\kappa M\ell }{4}=\sqrt{3}$) . \ For this range of
parameters, the candlestick gets a bulge near $t=0$ due to the presence of
the point mass. \bigskip 
For $\xi <0$ this effect does not occur, except that the first-derivative of
the evolution of the circumference becomes discontinuous at $t=0$. \ 

\begin{figure}[tbp]
\begin{center}
\epsfig{file=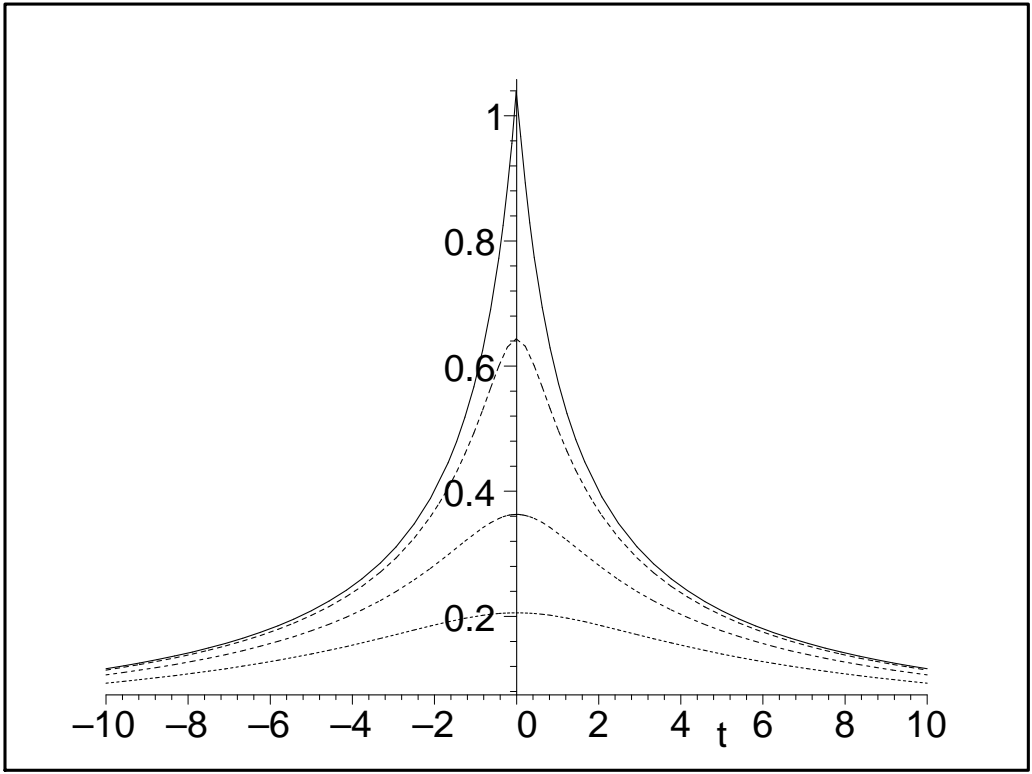,width=0.8\linewidth}
\end{center}
\caption{Circumference plotted against time as given in (\ref{an7},\ref{an7a}%
), with the choice $\protect\xi =2$ for differing values of ${\cal M}=%
\protect\sqrt{3}$(solid), $2$ (hash), $3$ (dash), $5$ (dot), with $\Lambda
>0 $. }
\label{fig17}
\end{figure}

Setting next $\tau =t/\ell ^{2}$ and $\left| \Lambda \right| =2/\ell ^{2}\,$%
\ yields from (\ref{gam1psol}) 
\begin{equation}
{\frak C}=\frac{2\ell }{\sqrt{\left( t/\ell \right) ^{2}-1}}\text{arctanh}%
\left( \frac{\xi \sqrt{\frac{\left( \kappa M\ell \right) ^{2}}{16}+(\xi
^{2}-1)(\left( t/\ell \right) ^{2}-1)}-\frac{\kappa M\ell }{4}}{\frac{\left(
\kappa M\ell \right) ^{2}}{16}+\xi ^{2}(\left( t/\ell \right) ^{2}-1)}\sqrt{%
\left( t/\ell \right) ^{2}-1}\right)  \label{an7}
\end{equation}%
for the proper circumference of the circle. The Hamiltonian is $H=\frac{2%
{\frak C}}{\kappa \ell ^{2}}$ and the spatial coordinate $x$ has been
rescaled as before. This solution matches analytically onto the $n=0$
solution of (\ref{gam1psoltrig}), which is \ 
\begin{equation}
{\frak C}=\frac{2\ell }{\sqrt{1-\left( t/\ell \right) ^{2}}}\left( \arctan
\left( \frac{\xi \sqrt{\frac{\left( \kappa M\ell \right) ^{2}}{16}+(1-\xi
^{2})(1-\left( t/\ell \right) ^{2})}-\frac{\kappa M\ell }{4}}{\frac{\left(
\kappa M\ell \right) ^{2}}{16}-\xi ^{2}(1-\left( t/\ell \right) ^{2})}\sqrt{%
1-\left( t/\ell \right) ^{2}}\right) +n\pi \right)  \label{an7a}
\end{equation}%
provided $\xi <\sqrt{1+\frac{\left( \kappa M\right) ^{2}}{8\Lambda }}$. In
this case the behaviour of the circumference is similar (though not
identical) to that described in eq. (\ref{an1}): it expands from zero at $%
t=-\infty $ to a maximum and back to zero again, with a cusp developing if
the left-hand bound of (\ref{xiconstraint}) is saturated, as shown in fig. \ %
\ref{fig17}. \bigskip 
For $n>0$ the behaviour of the spacetime is analogous to that described by
eq. (\ref{an10}) and fig. (\ref{fig16}): the circumference shrinks from
infinity beginning at $t=\ell $ down to a minimum, expands to a local
maximum \ at $t=0$ if (\ref{an11}) is satisfied, and then reverses its
evolution.

\begin{figure}[tbp]
\begin{center}
\epsfig{file=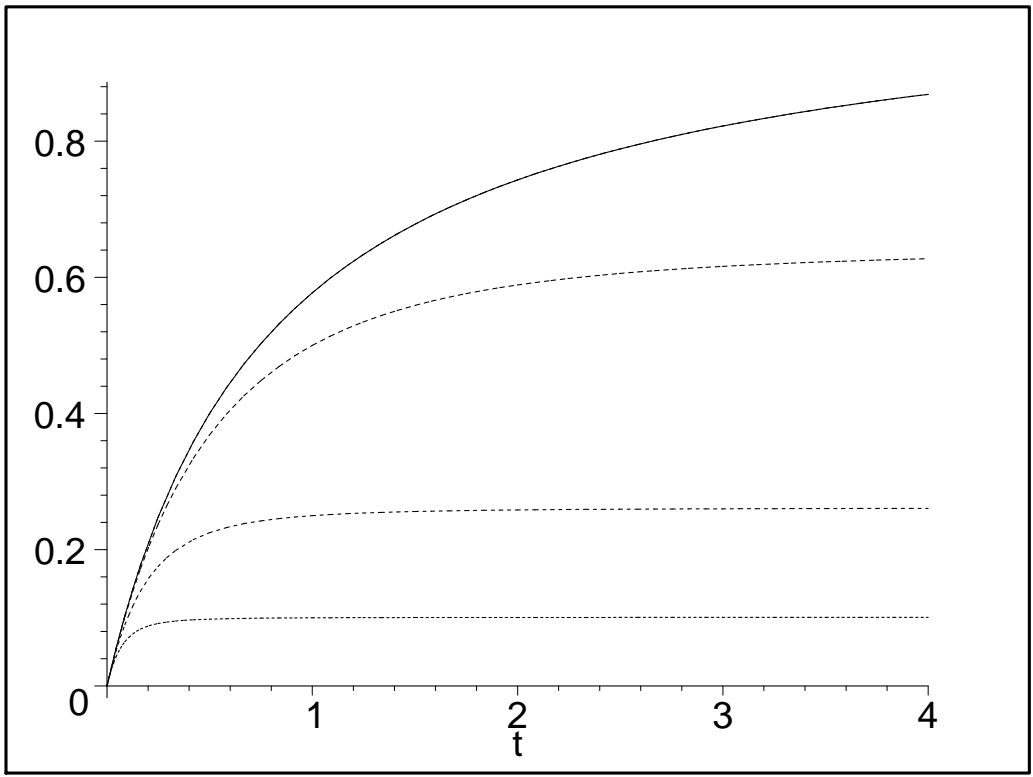,width=0.8\linewidth}
\end{center}
\caption{Circumference plotted against ``big bang'' time as given in (\ref%
{an8},\ref{an8a}), with the choice $\protect\xi =2$ for differing values of $%
{\cal M}=\protect\sqrt{3}$(solid), $2$ (hash), $4$ (dash), $10$ (dot), with $%
\Lambda >0$ and $n=0.$ }
\label{fig18}
\end{figure}

\begin{figure}[tbp]
\begin{center}
\epsfig{file=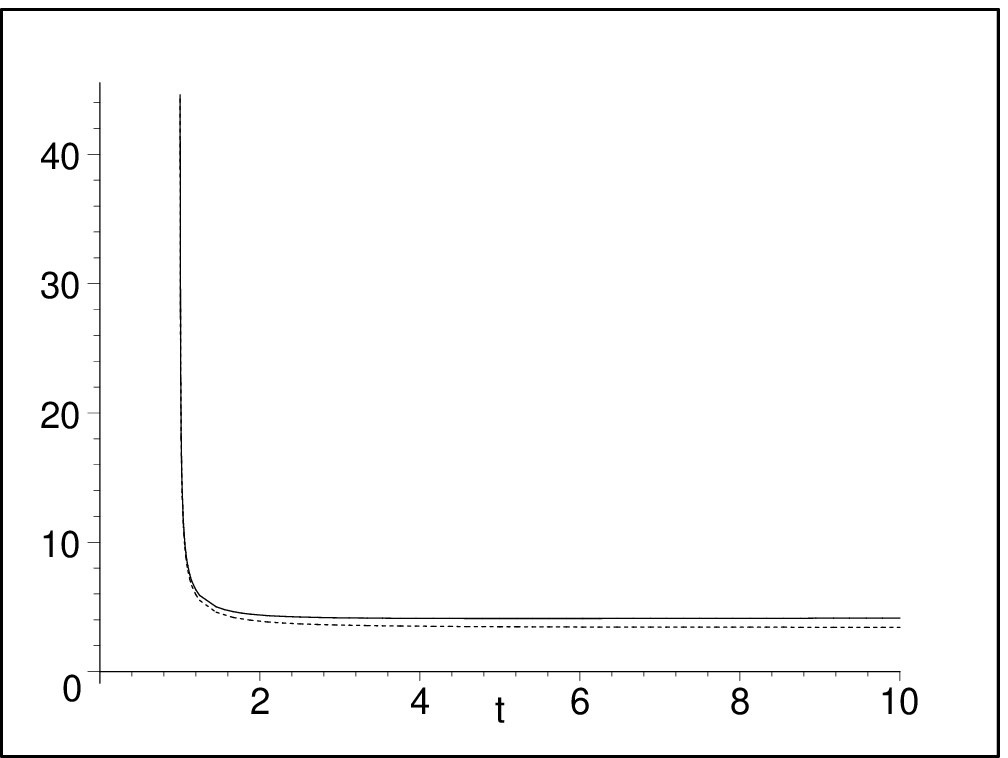,width=0.8\linewidth}
\end{center}
\caption{Circumference plotted against ``big bang'' time as given in (\ref%
{an8},\ref{an8a}), with the choice $\protect\xi =2$ for differing values of $%
{\cal M}=\protect\sqrt{3}$(solid) and $4$ (dash), with $\Lambda >0$ and $%
n=1. $ }
\label{fig19}
\end{figure}

For the``big bang'' coordinate choice, again replace\ $t\rightarrow \ell
^{2}/t$ \ in (\ref{an7},\ref{an7a}) to respectively obtain 
\begin{equation}
{\frak C}=\frac{2\left| t\right| }{\sqrt{1-\left( t/\ell \right) ^{2}}}\text{%
arctanh}\left( \frac{\xi \sqrt{\frac{\left( \kappa M\ell \right) ^{2}}{16}%
\left( t/\ell \right) ^{2}+(\xi ^{2}-1)(1-\left( t/\ell \right) ^{2})}-\frac{%
\kappa M\ell }{4}\left| t/\ell \right| }{\frac{\left( \kappa M\ell \right)
^{2}}{16}\left( t/\ell \right) ^{2}+\xi ^{2}(1-\left( t/\ell \right) ^{2})}%
\sqrt{1-\left( t/\ell \right) ^{2}}\right)  \label{an8}
\end{equation}%
\begin{equation}
{\frak C}=\frac{2\left| t\right| }{\sqrt{\left( t/\ell \right) ^{2}-1}}%
\left( \arctan \left( \frac{\xi \sqrt{\frac{\left( \kappa M\ell \right) ^{2}%
}{16}\left( t/\ell \right) ^{2}+(1-\xi ^{2})(\left( t/\ell \right) ^{2}-1)}-%
\frac{\kappa M\ell }{4}\left| t/\ell \right| }{\frac{\left( \kappa M\ell
\right) ^{2}}{16}\left( t/\ell \right) ^{2}-\xi ^{2}(\left( t/\ell \right)
^{2}-1)}\sqrt{\left( t/\ell \right) ^{2}-1}\right) +n\pi \right)
\label{an8a}
\end{equation}%
and these solutions analytically continue into each other for $n=0$. \ The
circumference evolves from zero size and asymptotes to its maximal size
given by the limit 
\begin{equation}
\lim_{t\rightarrow \infty }{\frak C}\left( n=0\right) =2\ell \left( \arctan
\left( \frac{\xi \sqrt{\frac{\left( \kappa M\ell \right) ^{2}}{16}+(1-\xi
^{2})}-\frac{\kappa M\ell }{4}}{\frac{\left( \kappa M\ell \right) ^{2}}{16}%
-\xi ^{2}}\right) +n\pi \right)  \label{an13}
\end{equation}%
with $n=0$, and is described by fig. \ref{fig18} \bigskip 
For $n>1$, the circumference contracts from infinity beginning at $t=1$, and
again asymptotes to the value (\ref{an13}), as shown in fig. \ref{fig19}$.$
\bigskip 
There is also a scenario which is the time-reversed version of this, and is
obtained from (\ref{an8a}) for negative values of $t$.

\begin{figure}[tbp]
\begin{center}
\epsfig{file=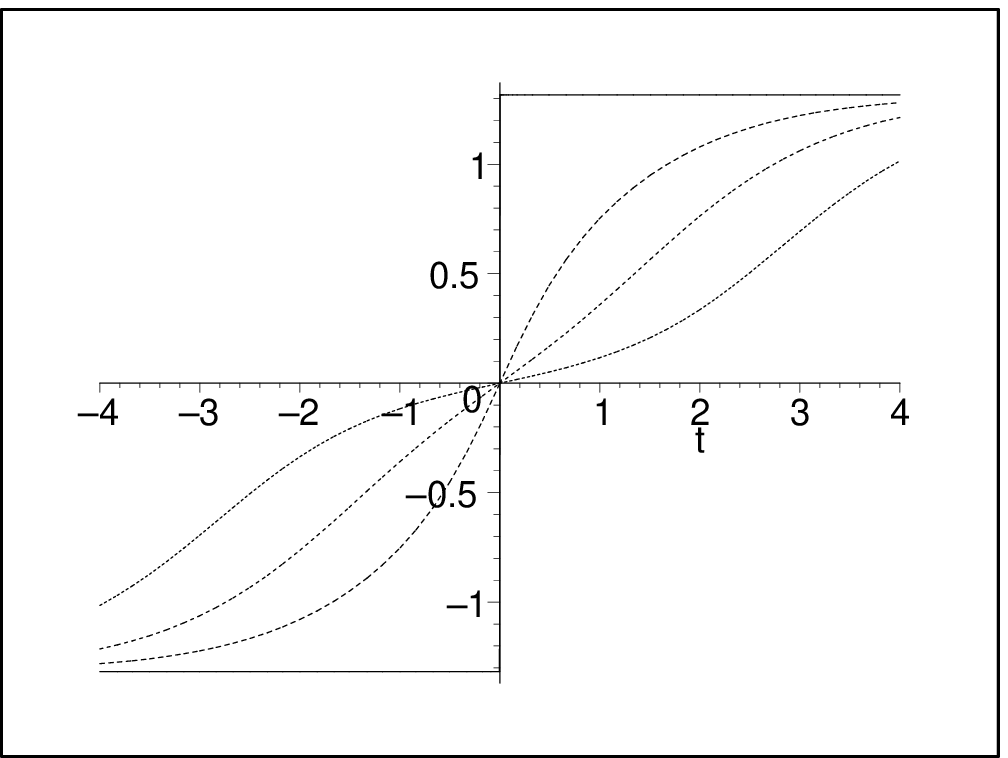,width=0.8\linewidth}
\end{center}
\caption{Hamiltonian associated with eq. (\ref{an14}) plotted against time
in conformally flat coordinates for$\ \protect\xi =2$ for differing values
of ${\cal M}=0$ (solid), $1$ (hash), $3$ (dash), $10$ (dot), with $\Lambda
>0 $.}
\label{fig20}
\end{figure}

Finally, consider the situation using conformal coordinates where $\dot{\tau}%
=\sqrt{\pm \left( \tau ^{2}-\Lambda /2\right) }$. \ Eqs. (\ref{gam1psol},\ref%
{gam1psoltrig}) respectively become 
\begin{equation}
{\frak C}=\frac{2\ell }{\sinh (t/\ell )}\text{arctanh}\left( \frac{\xi \sqrt{%
\frac{\kappa ^{2}M^{2}}{16}+(\xi ^{2}-1)\sinh ^{2}(t/\ell )}-\frac{\kappa M}{%
4}}{\frac{\kappa ^{2}M^{2}}{16}+\xi ^{2}\sinh ^{2}(t/\ell )}\sinh (t/\ell
)\right)  \label{an14}
\end{equation}%
\begin{equation}
{\frak C}=\frac{2\ell }{\left| \sin (t/\ell )\right| }\left[ \arctan \left( 
\frac{\xi \sqrt{\frac{\kappa ^{2}M^{2}}{16}+(1-\xi ^{2})\sin ^{2}(t/\ell )}-%
\frac{\kappa M}{4}}{\frac{\kappa ^{2}M^{2}}{16}-\xi ^{2}\sin ^{2}(t/\ell )}%
\left| \sin (t/\ell )\right| \right) +n\pi \right]  \label{an15}
\end{equation}%
with respective Hamiltonians $H=\frac{2{\frak C}}{\kappa t^{2}}\sinh (t/\ell
)$, $H=\frac{2{\frak C}}{\kappa t^{2}}\sin (t/\ell )$. \ When the
circumference is given by (\ref{an14}), it evolves from zero size to a
maximal value at $t=0$, and then reverses its evolution, similar to the
situation illustrated in fig. \ The corresponding value of the Hamiltonian
decreases from unity to zero when the maximal expansion is attained, and
then increases back to unity from zero again, as shown in fig. \ref{fig20}.
\bigskip 
The negative values of the Hamiltonian arise because of the choice of \ time
coordinate; \ $\tau $ is decreasing as the circumference is increasing in
the first half of the evolution. \ 

\begin{figure}[tbp]
\begin{center}
\epsfig{file=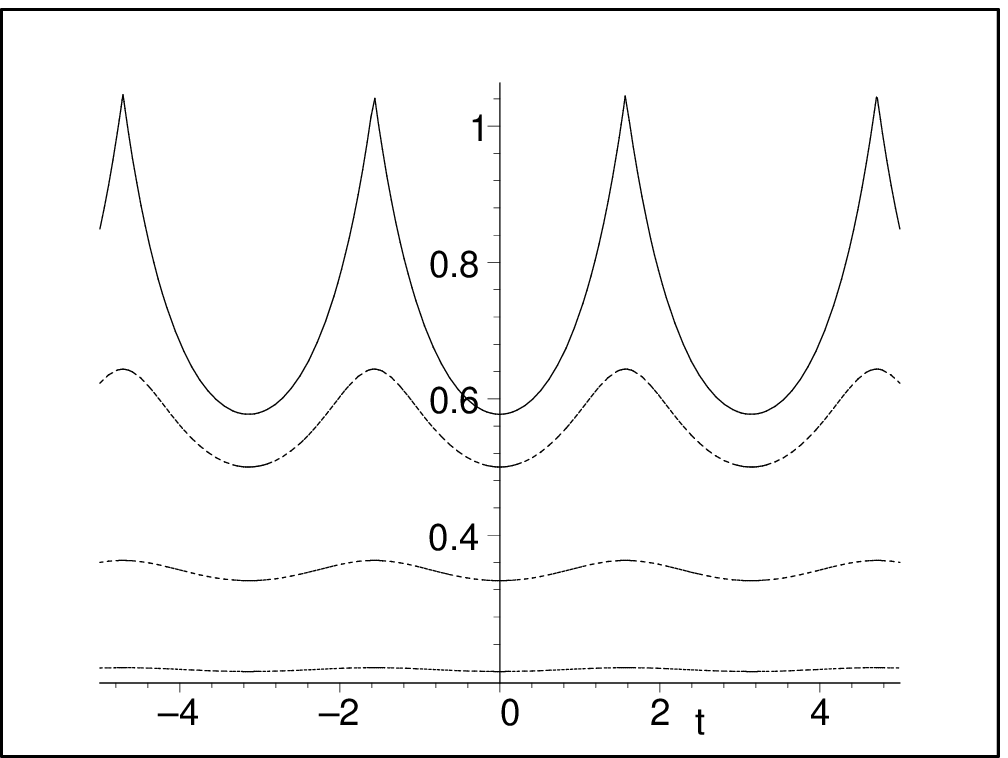,width=0.8\linewidth}
\end{center}
\caption{Circumference given in (\ref{an15}) plotted against time with the
choice $\protect\xi =2$ for differing values of ${\cal M}=\protect\sqrt{3}$%
(solid, with cusp), $2$ (hash), $4$ (dash), $10$ (dot), with $\Lambda >0$
and $n=0.$ The dashed line for ${\cal M}=4$ looks almost solid because of
the close proximity of the dash marks.}
\label{fig21}
\end{figure}

\begin{figure}[tbp]
\begin{center}
\epsfig{file=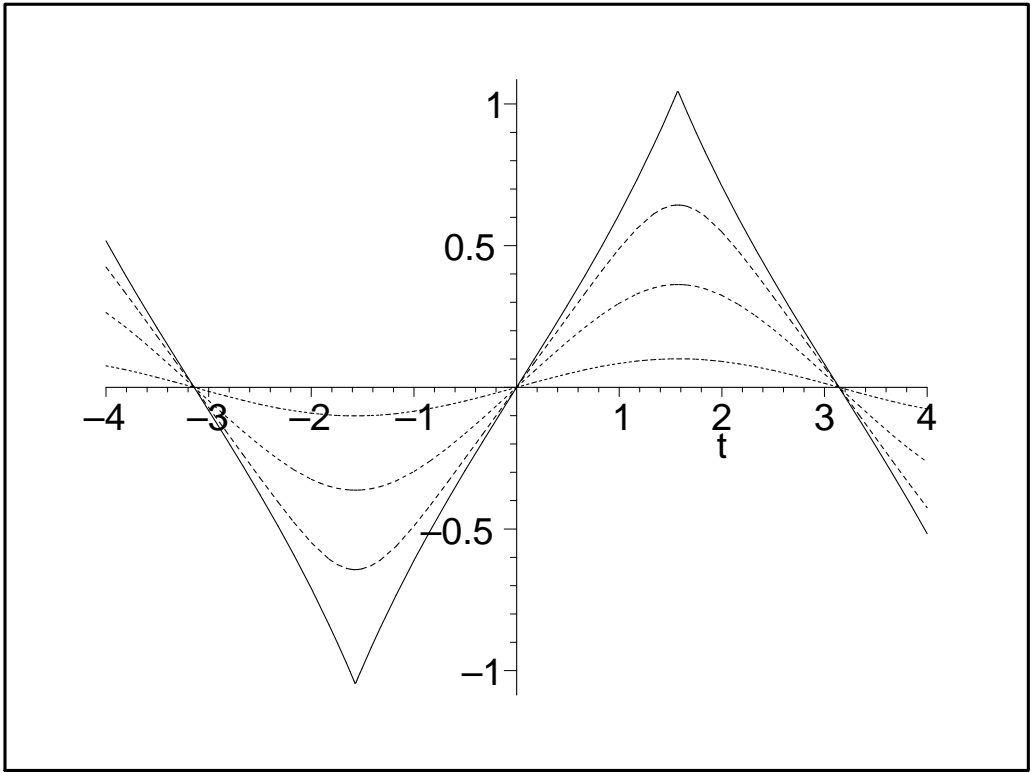,width=0.8\linewidth}
\end{center}
\caption{The corresponding Hamiltonian for the cases shown in (\ref{fig21})
plotted against time with the choice $\protect\xi =2$ for differing values
of ${\cal M}=\protect\sqrt{3}$(solid), $2$ (hash), $4$ (dash), $10$ (dot),
with $\Lambda >0$ and $n=0.$ }
\label{fig22}
\end{figure}

\begin{figure}[tbp]
\begin{center}
\epsfig{file=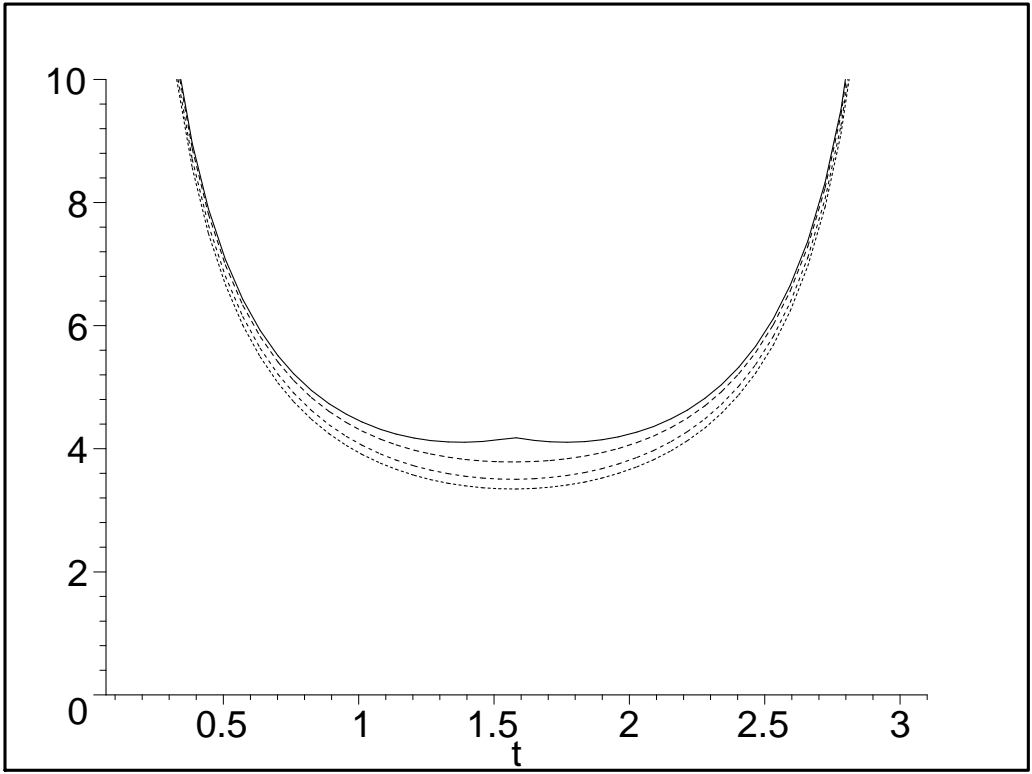,width=0.8\linewidth}
\end{center}
\caption{Circumference given in (\ref{an15}) plotted against time with the
choice $\protect\xi =2$ for differing values of ${\cal M}=\protect\sqrt{3}$%
(solid, with cusp), $2$ (hash), $4$ (dash), $10$ (dot), with $\Lambda >0$
and $n=1.$}
\label{fig23}
\end{figure}

\begin{figure}[tbp]
\begin{center}
\epsfig{file=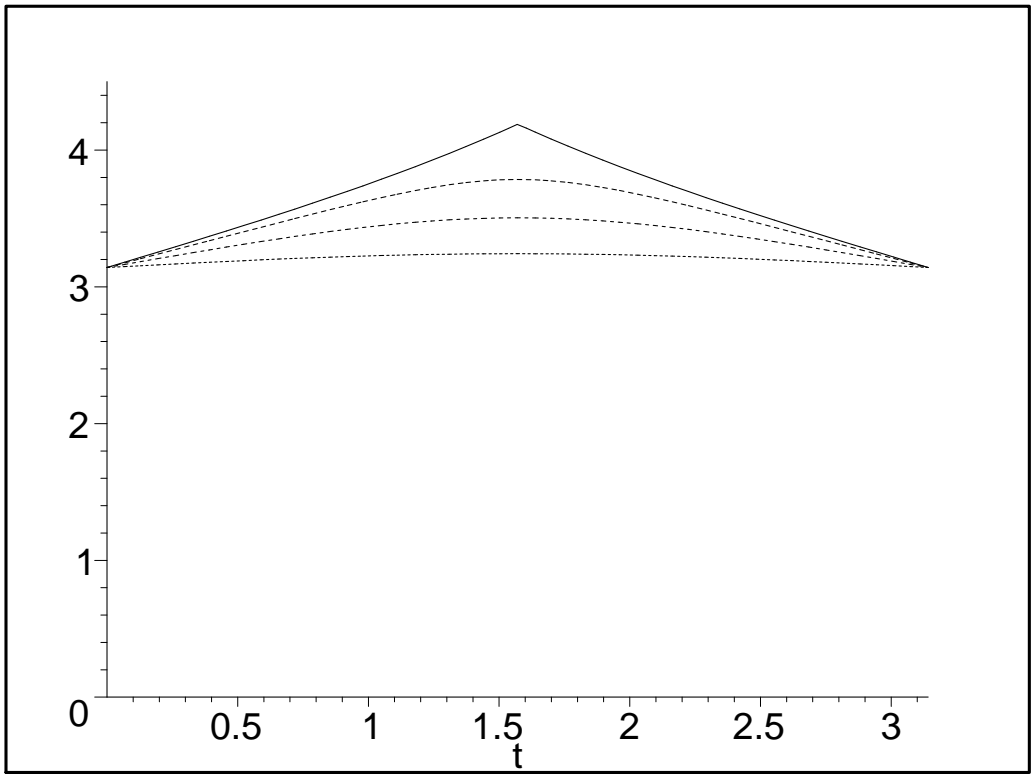,width=0.8\linewidth}
\end{center}
\caption{The corresponding Hamiltonian for the cases shown in (\ref{fig23})
plotted against time with the choice $\protect\xi =2$ for differing values
of ${\cal M}=\protect\sqrt{3}$(solid), $2$ (hash), $4$ (dash), $10$ (dot),
with $\Lambda >0$ and $n=1.$}
\label{fig24}
\end{figure}

When the circumference is given by (\ref{an15}) the situation is markedly
different. \ For $n=0$ it undergoes periodic oscillations between $\pm 2\ell
\arctan \left( \frac{\xi \sqrt{\frac{\kappa ^{2}M^{2}}{16}+(1-\xi ^{2})}-%
\frac{\kappa M}{4}}{\frac{\kappa ^{2}M^{2}}{16}-\xi ^{2}}\right) $, with
cusps developing whenever the left-hand side of eq.(\ref{an11}) is
saturated, as shown in fig.\ref{fig21}. \bigskip 
The accompanying Hamiltonian oscillates between positive and negative values
as shown in fig. \ref{fig22}. \bigskip 
For $n>1$, the circumference diverges at $t=0$, decreases to some minimal
value, and then expands out to infinity at $t=\pi \ell $, with the
Hamiltonian correspondingly increasing from $4\pi /\kappa \ell $ to a
maximum and then decreasing back to this value. If the inequality (\ref{an11}%
) is respected, then there will be a local maximum at $t=\pi /2$ in the
circumference, but not in the Hamiltonian. Figs. \ref{fig23} and \ref{fig24}
depict the behaviours. \bigskip 
\bigskip 
For $t\in \left[ 2n\pi \ell ,\left( 2n+1\right) \pi \ell \right] $ the
patterns shown in figs. \ref{fig23}, \ref{fig24}\ are the same. However for $%
t\in \left[ \left( 2n+1\right) \pi \ell ,2n\pi \ell \right] $ the
circumference in (\ref{an15}) becomes negative, and the solution is invalid.

\section{Summary}

\bigskip

The canonical formulation of lineal gravity on a circle yields a rich set of
interesting spacetime dynamics when coupled to point particles. \ The
explicit solutions obtained in this paper for the single-particle case
represent a new set of exact solutions to the equations of lineal gravity,
and illustrate the broad range of spacetime behaviours that can arise for a
self-gravitating system in a compact spacetime.

The formalism developed in this paper could be used to treat a variety of
related problems, including solving the 2-body problem (with and without
charge), generalization to other dilatonic theories of gravity,\ exploring
new solutions to the static balance problem (extending the work of ref. \cite%
{statbal}), and developing a statistical mechanics for self-gravitating
systems in compact spacetimes. \ 

A very interesting open problem is the quantization of the system described
here. \ The lineal gravity system described here has $(N-1)$ degrees of
freedom, which is zero for a single point particle. \ The nonlinearities of
the system here are much less severe than in $(3+1)$ dimensional gravity,
and so the prospects for achieving this goal do not seem too remote. Work on
this area is in progress.

\bigskip

\bigskip {\Large Acknowledgements}

I am grateful to V. Moncreif for interesting discussions concerning this
work, and to T. Ohta for helpful correspondence. I am grateful for the
hospitality of the Institute for Theoretical Physics at Santa Barbara, where
part of this work was carried out. This research was supported by the
Natural Sciences and Engineering Research Council of Canada.


\bigskip \bigskip

\begin{thebibliography}{99}
\bibitem{fractal} H. Koyama and T. Kinoshi, astro-ph/0008208

\bibitem{yawn} See B.N. Miller and P. Youngkins, Phys. Rev. Lett. {\bf 81}
4794 (1998); K.R. Yawn and B.N. Miller, Phys. Rev. Lett. {\bf 79} 3561
(1997) and references therein.

\bibitem{pchak} R.B. Mann and P. Chak, gr-qc/0101106

\bibitem{italiangroup} A. Bellini, M. Ciafaloni and P. Valtancoli, Nucl.
Phys. {\bf B462} (1996) 453; L. Cantini, P. Menotti, D. Seminara,
hep-th/0012022.

\bibitem{OR} T. Ohta and R.B. Mann, Class. Quant. Grav. {\bf 13} (1996) 2585.

\bibitem{2bd} R.B. Mann and T. Ohta, Phys. Rev. {\bf D57} (1997) 4723;
Class. Quant. Grav. {\bf 14} (1997) 1259.

\bibitem{2bdcossh} R.B. Mann, D. Robbins and T. Ohta, Phys. Rev. Lett. {\bf %
82} (1999) 3738.

\bibitem{2bdcoslo} R.B. Mann, D. Robbins and T. Ohta, Phys. Rev. {\bf D60}
(1999) 104048.

\bibitem{2bdchglo} R.B. Mann, D. Robbins, T. Ohta and M.\ Trott, Nucl. Phys. 
{\bf B590} \ 367.

\bibitem{marco} R.B. Mann, G. Potvin and M. Raiteri, Class.\ Quant. Grav. 
{\bf 17} (2000) 4941.

\bibitem{Moncreif} V. Moncreif, J.\ Math. Phys. {\bf 30} (1989) 2907.

\bibitem{r3} R.B. Mann, Found. Phys. Lett. {\bf 4} (1991) 425; R.B. Mann,
Gen. Rel. Grav. {\bf 24} (1992) 433.

\bibitem{jchan} S.F.J. Chan and R.B. Mann, Class. Quant. Grav. {\bf 12}
(1995) 351.

\bibitem{JT} R. Jackiw, Nucl. Phys. B {\bf 252}, 343 (1985); C. Teitelboim,
Phys. Lett. B {\bf 126}, 41, (1983).

\bibitem{BanksMann} T. Banks and M. O' Loughlin, Nucl. Phys. {\bf B362 }
(1991) 649; R.B. Mann, Phys. Rev.{\bf D47} (1993) 4438.

\bibitem{admV} R. Arnowitt, S. Deser and C.W. Misner, Phys. Rev {\bf 120}
(1960) 313.

\bibitem{statbal} R.B. Mann and T. Ohta, Class.\ Quant. Grav. {\bf 17}
(2000) 4059.
\end{thebibliography}
\end{document}